\documentclass[a4paper,11pt]{article}
\usepackage{amsmath,amssymb,ascmac,enumerate,booktabs,amsthm}
\usepackage{slashed}
\usepackage{braket}
\usepackage{url}
\usepackage{cite}
\usepackage{color}
 \usepackage{amsthm} 

\def\slashb#1{\not\!\!#1}
\usepackage[dvipdfmx]{hyperref}

\makeatletter
 
  \@addtoreset{equation}{section}
\makeatother

\begin{document}

\begin{titlepage}
\begin{flushright}
DIAS-STP-23-03 \\
\today
\end{flushright}
\vspace{0.5cm}
\begin{center}
{\Large \bf
Anomaly Inflow of Rarita-Schwinger Field \\
in $3$ Dimensions
}

\lineskip .75em
\vskip 2.5cm
{\large  Saki Koizumi \footnote{saki@stp.dias.ie}}
\vskip 2.5em
 {\normalsize\it School of Theoretical Physics,
Dublin Institute for Advanced Studies,\\
$10$ Burlington Road, Dublin $4$, Ireland}
\vskip 3.0em


\end{center}


\begin{abstract}
We study the anomaly inflow of the Rarita-Schwinger field with gauge symmetry in $3$ dimensions.
We find that global anomalies of the Rarita-Schwinger field are obtained by the spectral flow, which is similar to Witten's $SU(2)$ global anomaly for a Weyl fermion.
The Rarita-Schwinger operator is shown to be a self-adjoint Fredholm operator, and its spectral flow is determined by a path on the set of self-adjoint Fredholm operators with the gap topology.
From the spectral equivalence of the spectral flow, we find that the spectral flow of the Rarita-Schwinger operator is equivalent to that of the spin-$3/2$ Dirac operator.
From this fact, we confirm that the anomaly of the $3$-dimensional Rarita-Schwinger field is captured by the anomaly inflow.
Finally, we find that there are no global anomalies of gauge-diffeomorphism transformations on spin manifolds with any gauge group.
We also confirm that the anomalous phase of the partition function which corresponds to the generator of $\Omega_4^{{\rm Pin}^+}(pt)=\mathbb{Z}_{16}$ is $\exp(3i\pi /8)$ for the Rarita-Schwinger theory on unorientable ${\rm Pin}^+$ manifolds without gauge symmetry.
\end{abstract}

\end{titlepage}
\baselineskip=0.5cm

\tableofcontents

\section{Introduction}
Gauge anomalies are the obstruction to defining quantum field theories.
 Perturbative anomalies of matter theories have been studied extensively, which is an anomaly of a gauge transformation that connects to the trivial transformation continuously.
A gauge transformation that cannot connect trivial gauge symmetry continuously may also have an anomaly, called a global anomaly \cite{Old-SU(2),GG=Witten}.

Let us consider anomalies of a massless Dirac fermion system in $d$ dimensions.
The partition function of a massless spin-$1/2$ fermion on a $d$-dimensional manifold $X$ is the determinant or Phaffian of the spin-$1/2$ Dirac operator \cite{EW}.
The phase of the partition function is a section of the determinant line bundle \cite{Q,BF1,BF2,F1,F2}, which is a complex line bundle on the configuration space of the set of a gauge field and a metric on $X$.
Global anomalies are obtained by the spectral flow of the Dirac operator, which is related to the topological indices \cite{APS1,APS2,APS3} on the mapping torus \cite{TI1,TI2,TI3,EW,GG=Witten ,Old-SU(2)}.
On the other hand, the path integral of a massive Dirac fermion system on a $(d+1)$-dimensional bulk spacetime $Y$ with the chiral boundary condition \cite{WY} is exponential of the $\eta$-invariant of the Dirac operator on $Y$ with the Atiyah-Patodi-Singer (APS) boundary condition \cite{EW,KY,WY}.
This path integral on $Y$ is changed by a gauge-diffeomorphism transformation $\pi$ on the boundary by $\exp(i\pi\eta(M_\pi))$, and this change is equivalent to the global anomaly of the original theory \cite{KY,WY}, called the anomaly inflow.
Here, $M_\pi=X\times [0,1]$ is the mapping torus of a gauge-diffeomorphism $\pi$ and $Y+M_\pi$ is the connected sum of $Y$ and $M_\pi$.
Due to the Dai-Freed theorem \cite{Dai-Freed}, if a $(d+1)$-dimensional manifold $Y$ has a boundary $\partial Y=X$, the exponential $\eta$-invariant of the Dirac operator on $Y$ with the APS boundary condition \cite{APS1} is a section of the determinant bundle on $X$.
Therefore, the product of the determinant (or Phaffian) of the spin-$1/2$ Dirac operator on $X$ and the inverse of the path integral over a bulk manifold $Y$ is a complex constant on the configuration space of the set of a gauge field and a metric on $X$, which means the anomaly inflow \cite{KY,WY,EW}.

From the viewpoint of the anomaly inflow, we can identify the partition function of the original $d$-dimensional theory as the path integral on a bulk massive fermion system.
There is an ambiguity to choose a bulk manifold $Y$, where $\partial Y=X$.
The difference between two path integrals over two different choices of bulk manifolds $Y_1$ and $Y_2$ is given as
the $\eta$-invariant of the Dirac operator on $Y_1+\overline{Y}_2$, where $\overline{Y}_2$ is the orientation reversing of $Y_2$.
The global anomalies are classified by classes of the bordism group $\Omega_d^{\cal S}(BG)$ \cite{Kap14,Fre14,Garcia,KY-Cob,Kap14-1,FH-ref}, where $\Omega_d^{\cal S}(BG)$ is the $(d+1)$-dimensional bordism group with a structure ${\cal S}$ on the classification space $BG$ of a group $G$, which is the set of $(d+1)$-dimensional closed manifold with a ${\cal S}$-structure and $G$-bundle.

The anomaly inflow is generalized to other quantum field theories and string theories \cite{Garcia,p-form,FH-ref,FH-19,KY-YL}.
In particular, according to \cite{FH-ref,FH-19}, the partition function of the Rarita-Schwinger field \cite{RS} in $d$ dimensions is invertible \cite{Inv-F} and is the Phaffian of a spin-$3/2$ Dirac operator \cite{Freed-K,FM05}.
Furthermore, the corresponding anomalous theory \cite{FH-19} is studied.
The Rarita-Schwinger field is a spin-$3/2$ component field of the supergravity multiplet, and therefore it is important to study anomalies of theories that include gravity.

We consider the anomaly inflow of the $3$-dimensional Rarita-Schwinger theory in detail.
It was found that the perturbative anomalies of the Rarita-Schwinger field in even dimensions are equivalent to those of the spin-$3/2$ Dirac operator \cite{Witten-AG}.
In \cite{GG=Witten}, it was found that this equivalence is also valid for global anomalies.
Here, a spin-$3/2$ field is an element of the Hilbert space obtained by the completion of the set of sections on the tensor bundle of the tangent bundle and a spinor bundle, which is not necessary to satisfy the Rarita-Schwinger equation.
In this paper, we will show this equivalence of global anomalies by comparing the spectral flow of these two operators directly.

We will first see that the phase of the partition function of the Rarita-Schwinger field on a spin manifold is the $\eta$-invariant of the Rarita-Schwinger operator, and the global anomalies are determined by the spectral flow of the Rarita-Schwinger operator.
Here, the spectral flow is determined by a path on the set of self-adjoint Fredholm operators with the gap topology.
In this paper, we define the Rarita-Schwinger operator as a self-adjoint Fredholm operator which appears in the gauge-fixed Rarita-Schwinger effective action minus the effective action of the ghost field.
To compare the spectral flow of the Rarita-Schwinger operator and the spin-$3/2$ Dirac operator, we consider two paths on the set of self-adjoint Fredholm operators which correspond to a gauge-diffeomorphism transformation.
It is found that the spectral flow is homotopy invariant \cite{unbound-F}.
We construct a loop on the set of self-adjoint Fredholm operators, which connects those two paths.
Then, we find that the spectral flow of the Rarita-Schwinger operator and the spin-$3/2$ Dirac operator is equivalent.
Finally, we will find that the partition function of this theory is equivalent to the path integral of the massive spin-$3/2$ Dirac field and two massive bosonic spin-$1/2$ Dirac fermions on $4$-dimensional bulk manifold with a boundary condition.
We will also consider the case on an unorientable ${\rm Pin}^+$ manifold.
By using the anomaly inflow, we can identify the global anomaly of a gauge-diffeomorphism of the original $3$-dimensional Rarita-Schwinger field as a bordism class of the $4$-dimensional bordism group.
We will consider the classification of global anomalies by determining the anomalous phase for each bordism class.

The rest of this paper is organized as follows. 
In section $2$, we briefly review the anomaly inflow of the spin-$1/2$ Dirac fermion based on \cite{EW, KY, WY}, which is useful to consider the anomaly inflow of the Rarita-Schwinger field.
In section $3$, we consider the anomaly inflow of the Rarita-Schwinger field in $ 3$ dimensions.
We first show that the partition function of the Rarita-Schwinger field is given by the exponential of the $\eta$-invariant of the Rarita-Schiwnger operator.
From \cite{APS3}, we can see that the global anomaly is given by the spectral flow of the Rarita-Schwigner field, which is defined as a path on the set of self-adjoint Fredholm operators.
We finally find that the bulk theory is a massive spin-$3/2$ Dirac fermion in the meaning of the anomaly inflow.
In section $4$, we find that there are no global anomalies of gauge-diffeomorphism transformations on any $3$-dimensional spin manifolds.
 We also consider the classification of global anomalies for the case on a ${\rm Pin}^+$ manifold without gauge symmetry.
 In Appendix A, we explain the $1.5$ order formalism, which is important to consider the effective action of the Rarita-Schwinger theory based on \cite{Freedman-text}.
 In Appendix B, we will review the spectral flow of the Fredholm operators, which we used in section $3$ to compare the global anomaly of the Rarita-Schwinger field and that of a spin-$3/2$ Dirac fermion.

\section{Anomaly Inflow of Massless Dirac Fermion}
In this section, we begin with reviewing the anomaly inflow for a massless Dirac fermion system on a $3$-dimensional manifold based on \cite{EW,KY,WY}.
Since a spin-$3/2$ field is a section on the tensor product of the spinor bundle and the tangent bundle, we will use the idea of the anomaly inflow of a spin-$1/2$ Dirac fermion.

\subsection{Spin Manifold}
Let $X$ be a $3$-dimensional spin manifold with Lorentz signature.
We consider a real fermion $\psi$ which is defined as a section of a tensor bundle of a spinor bundle and a $G$-bundle on $X$ with a real representation of $G$.
Here, $G$ is a Lie group.
We denote $x^\mu$ the local coordinates of $X$, where $\mu=0$ is the time direction and $\mu=1,2$ the space directions, and denote $g_{\mu\nu}^L$ the metric on the manifold $X$.
We define the frame bundle with the frame fields $e^a(x)$ ($a=0,1,2$).
We introduce the vielbein $e^a{}_\mu(x)$ by $e^a(x)=e^a{}_\mu(x) dx^\mu$.
According to $SO(2,1)$ transformation on the frame bundle, a spinor field on $X$ is transformed by the spinor representation of ${\rm Spin}(2,1)$.
The generators of ${\rm Spin}(2,1)$ is constructed by the generators of the Clifford algebra $\gamma_\mu$ which satisfies $\{\gamma_\mu,\gamma_\nu\}=2g_{\mu\nu}^L$.
In the spinor representation, the generators of the Clifford algebra are given as $\gamma_\mu:=e_\mu^a\sigma^L_a$.
Here, $\sigma_a^L$, $a=0,1,2$ are defined as 
$\sigma_0^L:=i\sigma_2$,
$\sigma_1^L:=\sigma_1$,
$\sigma_2^L:=\sigma_3$, where $\sigma_a$ ($a=1,2,3$) are the Pauli matrices:
\begin{align}
\sigma_1=\left(
\begin{array}{cc}
0&1\\
1&0
\end{array}
\right),
\qquad
\sigma_2=\left(
\begin{array}{cc}
0&-i\\
i&0
\end{array}
\right),
\qquad
\sigma_3=
\left(
\begin{array}{cc}
1&0\\
0&-1
\end{array}
\right).\label{Pauli}
\end{align}
We now study the massless Dirac fermion system on the manifold $X$ with the Euclidian metric $g_{\mu\nu}$ by performing the Wick rotation $x^0\to x^0_E=ix^0$, $x^j\to x^j_E=x^j$ $(j\neq 0)$.
The corresponding gamma matrices are defined by $\gamma_E^0=i\gamma^0$ and $\gamma_E^j=\gamma^j$ $(j\neq 0)$.
Then, all components of $\gamma_E^0$ are pure imaginary, and all components of $\gamma_E^j$ are real.
We assume that $X$ is compact in the Euclidian signature.
We regard $\psi$ to be a fermion as an element of $L^2(\mathbb{S}\otimes (P\times_GV))$, where $\mathbb{S}\to X$ is a spinor bundle, $P \times_GV\to X$ is a $G$-bundle with a real representation of a group $G$ on linear space $V$, 
and $L^2(\mathbb{S}\otimes (P \times_G V))$ is a Hilbert space which is defined by the completion of the set of sections on $\mathbb{S}\otimes (P \times_G V)\to X$ (See Appendix B).
Here, the Lebesgue inner product of two fermion fields $\psi_1$ and $\psi_2$ is defined by
\begin{align}
\left<\psi_1,\psi_2\right>:=\int_X d^3x\sqrt{g} \psi_1^\dag\psi_2.\label{norm-1}
\end{align}
We define the charge conjugation matrix by \cite{GG=Witten,EW}:
\begin{align}
{\rm C}:=\ast(i\sigma_2).
\label{C1}
\end{align}
Here, we denote $\ast$ as the complex conjugation operator.
${\rm C}$ satisfies 
\begin{align}
[{\rm C},i\slashed{D}_{1/2}]=0,\qquad
{\rm C}^2=-1.
\label{C2}
\end{align}
Here, $i\slashed{D}_{1/2}$ is the spin-$1/2$ Dirac operator which acts on $L^2(\mathbb{S}\otimes (P \times_G V))$ as an elliptic operator.
The representation of $\gamma^\mu_E$ Euclidian spacetime is pseudo-real.
\footnote{
Let consider a representation $\rho$ of a group $G$ on a vector space $V$.
We define $\overline{V}:=\{v^\ast|v\in V\}$.
When $V\simeq \overline{V}$ and there exist a unitary transformation $U:V\to\overline{V}$ that satisfies $\overline{\rho}(g)=U\rho(g) U^{-1}$ for all $g\in G$, we define $\tau:=U\ast$.
 Then, $\tau$ is anti-unitary and satisfies $\tau\rho(g)=\rho(g)\tau$.
By the Schur Lemma, there exists a complex number $c\in\mathbb{C}$ that satisfies $\tau^2=c$, and we find $c\tau=(\tau^2)\tau=\tau(\tau^2)=\tau c$ with $c=\pm 1$.
We say $(V,\rho)$ is pseudo-real if $\tau^2=-1$.
}
Since $\left<{\rm C}\psi_1,{\rm C}\psi_2\right>=\left<\psi_2,\psi_1\right>$ and $\left<\psi,{\rm C}\psi\right>=\left<{\rm C}^2\psi,{\rm C}\psi\right>=-\left<\psi,{\rm C}\psi\right>$, we find that $\psi$ is orthogonal to ${\rm C}\psi$ for arbitrary fermion $\psi$:
\begin{align}
\left<\psi,{\rm C}\psi\right>=0.\label{norm-2}
\end{align}
The action on $X$ with the Euclidian signature can be written as
\begin{align}
S_E=&\int_Xd^3x\frac{\sqrt{g}}{2}\left\{
\psi^T\sigma_2i\slashed{D}_{1/2}\psi
+({\rm C}\psi)^T\sigma_2i\slashed{D}_{1/2}({\rm C}\psi)
\right\}
.\label{maj-act}
\end{align}
Since the Dirac operator $i\slashed{D}_{1/2}$ is self-adjoint with respect to the inner product (\ref{norm-1}), we choose the complex basis of the spinor space as the set of all eigenmodes of the Dirac operator.
Since we have considered a real fermion system in the Lorentz signature, we will define the partition function in the Euclidean signature by the contribution from one of each pair of Kramers doubling, $(\psi_j,{\rm C}\psi_j)$ \cite{EW}, where $i\slashed{D}_{1/2}\psi_j=\lambda_j\psi_j$, $\lambda_j\in\mathbb{R}$.
Using the Pauli-Villars regularization, we obtain the partition function \cite{EW}:
\begin{align}
Z_\psi=\exp(i\pi\eta(i\slashed{D}_{1/2}(A,g))/2).
\label{spin-1/2-pf}
\end{align}
Here, we denote $\eta(i\slashed{D}_{1/2}(A,g))$ the reduced $\eta$-invariant (See eq.(\ref{eta=def}) for its definitions) of the spin-$1/2$ Dirac operator $i\slashed{D}_{1/2}(A,g)$ on $L^2(\mathbb{S}\otimes (P \times_G V))$, where $A$ is a gauge field and $g_{\mu\nu}$ is a metric on a manifold $X$.
Since the Pauli-Villars regularization does not break any symmetry of the theory, there are no perturbative anomalies.
By using eq.(\ref{dif-eta-1}), the reduced $\eta$-invariant is changed by a gauge-diffeomorphism transformation $\pi$ as
\begin{align}
\eta(i\slashed{D}_{1/2}(A^\pi,g^\pi))-\eta(i\slashed{D}_{1/2}(A,g))={\rm SF}(f_{D\to D_\pi}).\label{dif-eta-2}
\end{align}
The partition function eq.(\ref{spin-1/2-pf}) is therefore changed as $Z_\psi\to(-1)^{{\rm SF}(f_{D\to D_\pi})}Z_\psi$ by $\pi$.
Here, $A^\pi$ and $g^\pi$ the gauge field and metric after $\pi$-transformation, ${\rm SF}(f_{D\to D_\pi})$ is the spectral number of a path $f_{D\to D_\pi}$.
$f_{D\to D_\pi}$ a continuous path on the set of the elliptic self-adjoint operators ${\cal CF}^{sa}(L^2(\mathbb{S}\otimes (P \times_G V)))$ defined by (\ref{fredholm}) with the gap metric:
\begin{align}
f_{D\to D_\pi}:[0,1]\to{\cal CF}^{sa}(L^2(\mathbb{S}\otimes (P \times_G V)))
,\qquad t\mapsto i\slashed{D}(A_t,g_t).
\label{flow-path-1}
\end{align}
Here, $A_t$ and $g_t$ are a gauge field and a metric on $X$, which are given as
\begin{align}
A_t(x):=(1-t)A(x)+tA^\pi(x),\quad
g_t(x):=(1-t)g(x)+tg^\pi(x),\quad x\in X,\quad t\in[0,1].\label{flow-path-2}
\end{align}
The spectral flow number ${\rm SF}(f_{D\to D_\pi})$ is equal to the index ${\cal I}_{1/2}(\pi):=n_+-n_-$ on the mapping torus $M_\pi$ of $\pi$ \cite{APS3,EW}.
Here, the mapping torus $M_\pi$ is a $4$-dimensional closed manifold $S^1\times X$ with the gauge field and metric which are given as (\ref{flow-path-2}), where $t\in [0,1]$ is the $S^1$-direction with $t=0$ and $t=1$ are identified by the transition function $\pi$.
$n_\pm$ is the number of zero modes with positive/negative chirality on the mapping torus.
The global anomaly of a gauge-diffeomorphism transformation $\pi$ of the partition function eq.(\ref{spin-1/2-pf}) is given by $(-1)^{{\cal I}_{1/2}(\pi)/2}$.
(From eq.(\ref{norm-2}), all eigenmodes appear as a pair. Therefore ${\rm SF}(f_{D\to D_\pi})$ and ${\cal I}_{1/2}(\pi)$ are even numbers.)

We can identify the partition function eq.(\ref{spin-1/2-pf}) as the path integral over a bulk massive Dirac fermion system on a $4$-dimensional spin  manifold $Y$ with a boundary $\partial Y=X$ \cite{WY}, where Lagrangian is defined by
\begin{align}
{\cal L}:=\overline{\Psi}(i\slashed{D}_{1/2}-iM)\Psi
+\overline{{\rm C}\Psi}(i\slashed{D}_{1/2}-iM){\rm C}\Psi,\qquad\overline{\Psi}:=\Psi^T\Gamma^2,\qquad
M>0.
\end{align}
Here, the gamma matrices $\Gamma^\mu$ and the chiral matrix $\Gamma^5$ are given by
\begin{align}
\Gamma^0=&\left(
\begin{array}{cc}
0&\sigma_3\\
\sigma_3&0
\end{array}
\right),
\quad
\Gamma^1=\left(
\begin{array}{cc}
\sigma_1&0\\
0&\sigma_1
\end{array}
\right),
\quad
\Gamma^2=\left(
\begin{array}{cc}
\sigma_2&0\\
0&\sigma_2
\end{array}
\right),
\notag \\
\Gamma^3=&
\left(
\begin{array}{cc}
\sigma_3&0\\
0&-\sigma_3
\end{array}
\right),
\quad
\Gamma^5=
\left(
\begin{array}{cc}
0&-i\sigma_3\\
i\sigma_3&0
\end{array}
\right).
\label{4d-gamma}
\end{align}
We also define the charge conjugation ${\rm C}$ by
 \begin{align}
 {\rm C}:=\Gamma^2\ast.\label{C-4d}
 \end{align}
We assume that $Y$ is cylinder $X\times (0,1]$ near the boundary, and choose an elliptic boundary condition as \cite{WY}:
\begin{align}
L:(1-\Gamma^5)\Psi(x)=0,\qquad x\in X=\partial Y.\label{bc1/2-real}
\end{align}
Then, the original $3$-dimensional massless fermion system appears as localized modes on the boundary when we consider one of the contributions from ${\rm C}\Psi$ and $\Psi$.
The path integral of one of the fermions $\Psi,{\rm C}\Psi$ on $Y$ is \cite{WY}:
\begin{align}
Z(Y|L):=\exp(i\pi\eta(i\slashed{D}_{1/2}(Y|{\rm APS})))|Z(Y|L)|,\label{AF-1/2}
\end{align}
where $i\slashed{D}_{1/2}(Y|{\rm APS})$ is the Dirac operator on $Y$ with the global Atiyah-Patodi-Singer (APS) boundary condition.
By a gauge-diffeomorphism transformation $\pi$ on the original $3$-manifold $X$, the bulk is changed as $Y\to Y+M_\pi$, where $Y+M_\pi$ is the connected sum of two manifolds $Y$ and $M_\pi$ gluing the common boundary $X$ \cite{KY}.
By using the Dai-Freed theorem of the $\eta$-invariant with the APS boundary condition \cite{Dai-Freed}, the phase of the partition function eq.(\ref{AF-1/2}) is changed by $\exp(i\pi\eta(i\slashed{D}_{1/2}(M_\pi)))$.
Since $\{\Gamma^5,\slashed{D}_{1/2}\}=0$, all non-zero eigenvalues $\lambda$ of the Dirac operator appear as a pair $(\lambda,-\lambda)$.
Then, only zero modes contribute to the reduced $\eta$-invariant, and therefore $\eta(i\slashed{D}_{1/2}(M_\pi))$ is equivalent to half of the number of zero modes on $M_\pi$.
Since $[{\rm C},\Gamma^5]=0$, the number of positive/negative chirality zero modes on the mapping torus $n_\pm$ are both even.
Therefore, we find $\exp(i\pi\eta(M_\pi))=(-1)^{{\cal I}_{1/2}(\pi)/2}$.
This means that the global anomalies of the partition function eq.(\ref{spin-1/2-pf}) is equivalent to the change of the bulk $Y\to Y+M_\pi$, where $\partial Y=X$ and $M_\pi$ is the mapping torus.

In the case the complex representation of a gauge group, we cannot define real fermion both in the Euclidian and the Lorentzian signature.
Since the operator (\ref{C1}) does not commute with the gauge symmetry, eigenmodes do not appear as a pair $(\lambda,\lambda)$.
The partition function is then given as \cite{EW}:
\begin{align}
Z=\prod_j\lambda_j.
\end{align}
Here, $\prod_j$ is the product of all eigenmodes of the Dirac operator.
The global anomalies of a gauge-diffeomorphism $\pi$ is given as $(-1)^{{\cal I}_{1/2}(\pi)}$ \cite{EW}, where ${\cal I}_{1/2}(\pi)$ is the index of the Dirac operator on the mapping torus.
The anomalous phase of this partition function is equivalent to the path integral of a massive Dirac fermion system on a bulk spacetime with the chiral boundary condition \cite{WY}.

\subsection{${\rm Pin}^+$ Unorientable Manifold}
We consider the case where a $3$-dimensional spacetime $X$ is unorientable.
Any manifold $X$ has a canonical orientation double cover $\pi:\tilde{X}\to X$, which is a fiber bundle whose fiber at $x\in X$ is the set of orientations on $T_xX$.
We define a manifold:
\begin{align}
\hat{X}:=\{(x,\pi^{-1}(x))|\:x\in X\}.\label{DC}
\end{align}
This new manifold $\hat{X}$ is called the orientable double cover of $X$, which is an orientable manifold.
$\hat{X}$ is connected if $X$ is unorientable.
There exists an orientation reversing map $\tau:\hat{X}\to\hat{X}$ which satisfies $\tau^2=1$ and $X=\hat{X}/\tau$.
We consider the case $X$ has a ${\rm Pin}^+$ structure, which means the second Stiefel-Whitney class of $X$ satisfies $w_2(X)=0$ \cite{FH-19}.

Let us first consider the action of $\tau$ on a spinor on $\hat{X}$.
$\tau$ and $SO(3)$ act on the frame field as $O(3)$ transformations, and therefore it is the $O(3)$-transformation on the orthonormal frame bundle of the tangent bundle.
There is a short exact sequence:
\begin{align}
1\to\mathbb{Z}_2\to\underline{{\rm Pin}(3)}\xrightarrow{\widetilde{{\rm Ad}}}\underline{O(3)}\to 1.\label{UO-seq1}
\end{align}
Here, we use the notation $\underline{G}$ is the set of smooth maps from each open set of $\hat{X}$ to $G$, where $G$ is a Lie group.
$\widetilde{{\rm Ad}}:{\rm Pin}(3)\to O(3)$ is defined as follows:
\begin{align}
\left(\widetilde{{\rm Ad}}({\cal O}_P)V\right)^\mu:=\left\{
\begin{array}{ccc}
{\cal O}_P V^\mu {\cal O}_P^{-1},&\qquad&V\in Cl_3,\qquad {\cal O}_P\in {\rm Spin}(3),
\\
-{\cal O}_P V^\mu {\cal O}_P^{-1},&\qquad&V\in Cl_3,\qquad {\cal O}_P\in {\rm Pin}(3)\backslash{\rm Spin}(3),
\end{array}
\right.
\label{UO-1}
\end{align}
where $Cl_3$ is the $3$-dimensional real Clifford algebra.
By using the exact sequence eq.(\ref{UO-seq1}), we obtain the following cohomology long exact sequence:
\begin{align}
\ldots\xrightarrow{\delta^\ast} H^1(\hat{X},\mathbb{Z}_2)\to H^1(\hat{X},\underline{{\rm Pin}(3)})\xrightarrow{\widetilde{{\rm Ad}}}H^1(\hat{X},\underline{O(3)})\xrightarrow{\delta^\ast} H^2(\hat{X},\mathbb{Z}_2)\to\ldots.
\end{align}
Here, $\delta$ is the coboundary operator.
$H^1(\hat{X},\underline{{\rm Pin}(3)})$ and $H^1(\hat{X},\underline{O(3)})$ classify the principle ${\rm Pin}(3)$-bundles and the principle $O(3)$-bundles, whose elements correspond to the transition functions up to isomorphism of the principal bundles.
We denote $\pi_1:\mathbb{O}(\hat{X})\to\hat{X}$ the frame bundle of the tangentet bundle on $\hat{X}$, which is a principle $O(3)$-bundle on $\hat{X}$.
We consider local coordinates $\{U_\alpha\}_\alpha$ on $\hat{X}$, which is a local trivialization of $\pi_1:\mathbb{O}(\hat{X})\to\hat{X}$, and we denote $\{g_{\alpha\beta}\}_{\alpha,\beta}$ the set of the transition function of $\pi_1:\mathbb{O}(\hat{X})\to\hat{X}$.
Since $w_2(\hat{X}):=\delta^\ast(g_{\alpha\beta})=0$, there exists a principle ${\rm Pin}(3)$-bundle $\pi_2:\mathbb{P}{\rm in}(\hat{X})\to\hat{X}$ that satisfies $\widetilde{{\rm Ad}}(h_{\alpha\beta})=g_{\alpha\beta}$, where $h_{\alpha\beta}\in{\rm Pin}(3)$ is the transition function of a principle ${\rm Pin}(3)$-bundle, which satisfies the cocycle condition.
Then, there exist a bundle homomorphism $\Phi:\mathbb{P}{\rm in}(\hat{X})\to \mathbb{O}(\hat{X})$ that satisfies
$\pi_1(u)=\pi_2(\Phi(u))$ ($u\in{\rm O}(\hat{X})$) and $\Phi(ug)=\Phi(u)\widetilde{{\rm Ad}}(g)$ ($g\in{\rm Pin}(3)$).
This homomorphism $\Phi$ is the ${\rm Pin}^+$ structure.

Now, we determine the $O(3)$ action on the ${\rm Pin}^+(3)$-bundle.
It is enough to assume $\tau:x^\mu\to x^\mu-2w^\mu(x\cdot w)$, 
where $w=w^\mu \partial_\mu$ is a unit tangent vector.
\footnote{
Since we assume that $\hat{X}$ is a connected manifold, $\tau:\hat{X}\to\hat{X}$ reverses the orientation of $\hat{X}$ iff $\tau\circ i:\mathbb{R}^3\to\hat{X}$ and $i:\mathbb{R}^3\to\hat{X}$ is not isotopic.
Here, $i:\mathbb{R}^3\to\hat{X}$ is a map which is an isomorphism if we restrict the image of $i$ as $i:\mathbb{R}^3\to i(\mathbb{R}^3)\subset \hat{X}$.
}
By using eq.(\ref{UO-1}), we find that the representation ${\cal S}(\tau)$ of $\tau$ on the ${\rm Pin}^+(3)$-bundle is
\begin{align}
{\cal S}(\tau)\psi(x)=\pm w\cdot\gamma\psi(\tau x),\label{UO-4}
\end{align}
where $\psi$ is a section on the ${\rm Pin}^+$-bundle.
We also denote ${\cal S}(P)$ the representation of $P\in O(3)$ on the ${\rm Pin}^+(3)$-bundle.
Combining (\ref{UO-4}) and spinor representation of ${\rm Spin}(3)$, there are two possible transformation rules of a section $\psi$ of the ${\rm Pin}^+(3)$-bundle on $\hat{X}$ by any $P^\mu{}_\nu\in O(3)$ as \cite{EW}:
\begin{align}
\psi(x)\to \hat{S}(P)[\psi](x):=\left\{
\begin{array}{c}
{\cal S}(P)\psi(\tau x),
\\
{\rm det}\left({\cal S}(P)\right)\;{\cal S}(P)\psi(\tau x).
\end{array}
\right.\label{UO-3}
\end{align}
Here, $x\in\hat{X}$.
We denote ${\cal P}_\pm$ the ${\rm Pin}^+(3)$-bundle transformed by the first rule (second rule) of eq.(\ref{UO-3}), which relate two different irreducible representations of the ${\rm Pin}(3)$ group.

We will consider a Dirac fermion system on $X$.
We regard $\psi_\pm$ to be a fermion as an element of the Hilbert space $L^2({\cal P}_\pm\otimes (P \times_G V))$, where $P \times_G V\to \hat{X}$ is a $G$-bundle with a real representation of a Lie group $G$ on a linear space $V$.
The spin-$1/2$ Dirac operator $i\slashed{D}_{1/2}$ on $\hat{X}$ acts on $L^2(({\cal P}_+\oplus{\cal P}_-)\otimes (P \times_G V))$ anti-commutes with $\tau$ \cite{EW}:
\begin{align}
\{i\slashed{D}_{1/2},\tau\}=0.\label{anti-D}
\end{align}
Therefore, $i\slashed{D}_{1/2}$ is not a closed operator on $L^2({\cal P}_\pm\otimes (P \times_G V))$, but if we use the fact that $\psi^\dag_\pm\sigma_2\in L^2({\cal P}_\mp\otimes (P \times_G V))$ \cite{EW}, where $\psi_\pm\in L^2({\cal P}_\pm\otimes (P \times_G V))$, we construct the Dirac action only for the element of $L^2({\cal P}_+\otimes (P \times_G V))$ or $L^2({\cal P}_-\otimes (P \times_G V))$\cite{EW}:
\begin{align}
S_\pm=\int d^3x\sqrt{g}\overline{\psi}_\pm i\slashed{D}_{1/2}\psi_\pm,\qquad\overline{\psi}_\pm:=\psi_\pm^\dag\sigma_2,\qquad\psi_\pm\in L^2({\cal P}_\pm\otimes (P \times_G V)).\label{pin+}
\end{align}
Since $\{{\rm C},\tau\}=0$, we obtain ${\rm C}\psi_\pm\in L^2({\cal P}_\mp\otimes (P \times_G V))$, where $\psi_\pm\in L^2({\cal P}_\pm\otimes (P \times_G V))$ and ${\rm C}$ is defined in eq.(\ref{C1}).
Therefore, the partition function $Z_\pm$ of $L^2({\cal P}_\pm\otimes (P \times_G V))$ satisfies $Z_+=Z_-^\ast$.
The partition function of the action (\ref{pin+}) is given as the Phaffian of the spin-$1/2$ Dirac operator $i\slashed{D}_{1/2}(A,g)$ on $L^2({\cal P}_+\otimes (P \times_G V))$ with a gauge field $A$ and a metric $g$ \cite{EW}:
\begin{align}
Z_+={\rm Pf}(i\slashed{D}_{1/2}(A,g)).\label{pin+pf}
\end{align}
By a gauge-diffeomorphism transformation $\pi$, the set $(A,g)$ is changed to $(A^\pi,g^\pi)$, where $A^\pi$ and $g^\pi$ are a gauge field and a metric after the transformation.
Then, anomaly of $\pi$ is given as $Z_+\to (-1)^{{\rm SF}(f_{F\to D_\pi})}Z_+$, where $f_{D\to D_\pi}$ is given by eq.(\ref{flow-path-1}).

We can identify this partition function (\ref{pin+pf}) as the bulk path integral.
We consider a $4$-dimensional spin manifold $\hat{Y}$ with a boundary $\partial \hat{Y}=\hat{X}$, assuming that $\hat{Y}$ is a cylinder $\hat{X}\times I$ near the boundary.
We denote $Y=\hat{Y}/\tau_Y$, where $\tau_Y$ is an orientation reversing transformation which satisfies $\tau_Y^2=1$, and $\tau_Y=\tau$ near the boundary.
Then, $Y$ is a ${\rm Pin}^+$ manifold with a boundary $\partial Y=X$.
We can construct a ${\rm Pin}^+$ structure on $Y$.
But since there is only one irreducible representation of the ${\rm Pin}(4)$ group, we can define unique ${\rm Pin}^+(4)$ bundle on $Y$ as the associated vector bundle of a ${\rm Pin}^+$ structure.
We denote $\tilde{{\cal P}}_+\to\hat{Y}$ a ${\rm Pin}^+(4)$-bundle on $\hat{Y}$.
We also denote $\tilde{P}\times_GV\to\hat{Y}$ a $G$-bundle on $\hat{Y}$ with a real linear representation of $G$ on $V$.
We assume that $\tilde{{\cal P}}_+\to\hat{Y}$ and $\tilde{P} \times_G V\to\hat{Y}$ are the original ${\rm Pin}^+$ bundle and $G$-bundle on $3$-dimensional manifold on the boundary.
We consider a new Dirac operator on $L^2(\tilde{{\cal P}}_+\otimes(\tilde{P} \times_G V))$ \cite{EW}:
\begin{align}
\tilde{\slashed{D}}_{1/2}:=\Gamma^5\slashed{D}_{1/2}=Ui\slashed{D}_{1/2}U^{-1}=i\tilde{\Gamma}^\mu D_{1/2,\mu},\qquad U:=\frac{1-i\Gamma_5}{\sqrt{2}}.\label{DOPin4}
\end{align}
Here, $D_{1/2,\mu}$ the covariant derivative on spin-$1/2$ fields, $i\slashed{D}_{1/2}$ is the original $4$-dimensional spin-$1/2$ Dirac operator, and $\tilde{\Gamma}^\mu$ is defined by using eq.(\ref{4d-gamma}):
\begin{align}
\tilde{\Gamma}^\mu:=U\Gamma^\mu U^{-1}=-i\Gamma^5\Gamma^\mu.\label{New4d}
\end{align}
Since $\{\tau_Y,\Gamma^5\}=0$, we obtain
\begin{align}
[\tilde{\slashed{D}}_{1/2},\tau_Y]=0.\label{SA-4D}
\end{align}
Therefore, $i\tilde{\slashed{D}}_{1/2}$ is self-adjoint on $L^2(\tilde{{\cal P}}_+\otimes (\tilde{P} \times_G V))$.
We define ${\rm C}:=\ast\Gamma^2\Gamma^5=\ast(i\tilde{\Gamma}^2)$, where $\ast$ is the complex conjugate operator.
Then, ${\rm C}$ commute with any gauge transformation with a real representation and Lorentz transformations.
We find
\begin{align}
{\rm C}^2=-1,\qquad
[{\rm C},\tilde{\slashed{D}}_{1/2}]=0,\qquad
[{\rm C},\tau_Y]=0.\label{PR-4d}
\end{align}
Therefore, all eigenmodes of the Dirac operator on ${\rm Pin}^+(4)$ structure appear as a pair $(\lambda,\lambda)$.
We consider a massive Dirac fermion system on $Y=\hat{Y}/\tau_Y$ with a ${\rm Pin}^+(4)$ structure:
\begin{align}
S_4=\int_{\hat{Y}} d^4x\sqrt{g}\overline{\Psi}(i\tilde{\slashed{D}}_{1/2}+iM)\Psi,\qquad\overline{\Psi}:=\Psi^T\tilde{\Gamma}^2.
\end{align}
Here, $\Psi\in L^2(\tilde{{\cal P}}_+\otimes (\tilde{P} \times_G V))$.
We choose a boundary condition of the fermion as follows \cite{WY}:
\begin{align}
L:(1-\tilde{\Gamma}^0)\Psi|_{\partial Y=X}=0.\label{UO-bc}
\end{align}
Then, if we consider only one contribution from $\Psi$ and ${\rm C}\Psi$, the original massless theory defined by ${\cal P}_+$ appears as a localized mode on $X$.
The path integral of this bulk system is given as
\begin{align}
Z(L|Y)=\exp(i\pi\eta(i\tilde{\slashed{D}}_{1/2}(Y|{\rm APS})))|Z(Y|L)|.\label{Bp-U1/2}
\end{align}
Here, $i\tilde{\slashed{D}}_{1/2}(Y|{\rm APS})$ is the spin-$1/2$ Dirac operator on $L^2(\tilde{{\cal P}}_+\otimes (\tilde{P} \times_G V))$ with the APS boundary condition \cite{APS3}.
This phase of the path integral of this bulk system is changed by a gauge-diffeomorphism $\pi$ on the original manifold by $\exp(i\pi\eta(i\tilde{\slashed{D}}_{1/2}(M_\pi)))$, where $M_\pi$ is the mapping torus of $\pi$.
We also find $\exp(i\pi\eta\tilde{\slashed{D}}_{1/2}(M_\pi))=\exp(i\pi\eta\slashed{D}_{1/2}(M_\pi))$, where $i\slashed{D}_{1/2}(M_\pi)$ is the original spin-$1/2$ Dirac operator on the mapping torus $M_\pi$.
Therefore, we find that the partition function (\ref{pin+pf}) is equivalent to the path integral (\ref{Bp-U1/2}) in the same way as the case on a spin manifold, which is the anomaly inflow.
As shown in \cite{EW, Dai-Freed}, Phaffian of the Dirac operator on the boundary ${\rm Pf}(i\slashed{D}_{1/2}(A,g))$ and $\exp(i\pi\eta(i\tilde{\slashed{D}}_{1/2}(Y|{\rm APS})))$ are sections of the Phaffian line bundle on $X$.
Therefore, the product ${\rm Pf}(i\slashed{D}_{1/2}(A,g))\exp(-i\pi\eta(i\tilde{\slashed{D}}_{1/2}(Y|{\rm APS})))$ is complex constant on the configuration space of a gauge-field and a metric on $X$, which means global anomaly of the partition function (\ref{pin+pf}) is identified as the change of the path integral over the bulk spacetime by any gauge-diffeomorphism transformation on $X$.

\section{Anomaly Inflow of Rarita-Schwinger Field}
We will generalize the idea of anomaly inflow to the Rarita-Schwinger field \cite{RS} in $3$-dimensions.
In section 3.1, we will consider the anomaly inflow of the Rarita-Schwinger field on a $3$-dimensional spin manifold with a gauge group $G$ with a real representation.
In section 3.2, we will consider the case on an unorientable ${\rm Pin}^+(3)$ manifold.

\subsection{Spin Manifold}
Let $X$ be a $3$-dimensional compact spin manifold with Euclidian signature.
We regard $\psi^\mu$ to be a real spin-$3/2$ fermion as an element of the Hilbert space $L^2(\mathbb{S}\otimes TM\otimes (P \times_G V))$ (See Appendix B), where $\mathbb{S}\to X$ is a spinor bundle, $P \times_G V\to X$ is a $G$-bundle on $X$ with a real representation of a Lie group $G$ on a linear space $V$, 
 and $TX\to X$ is the tangent bundle.
Here, the Lebesgue inner product on spin-$3/2$ fields $\psi$ and $\chi$ is given as
\begin{align}
\left<\psi,\chi\right>:=\int_X d^3x\sqrt{g}\psi^\dag_\mu\chi^\mu.\label{inner-RS}
\end{align}
The action of the supergravity theory on $X$ with the Euclidian signature is given as follows \cite{RS}:
\begin{align}
S^E_{\rm SUGRA}=&S^E_2+S^E_{3/2},
\notag \\
S^E_2=&-\frac{1}{2\kappa^2}\int d^3x\:  \sqrt{g}\:e^{a\mu}e^{b\nu}R_{\mu\nu ab},
\notag \\
S^E_{3/2}=&\frac{1}{2}\int d^3x\:  \sqrt{g}\:\overline{\psi}_\mu\gamma^{\mu\nu\rho}(D_{3/2,\nu}\psi)_\rho
+\frac{1}{2}\int d^3x\:  \sqrt{g}\:\overline{{\rm C}\psi}_\mu\gamma^{\mu\nu\rho}(D_{3/2,\nu}{\rm C}\psi)_\rho.
\label{SG=Action}
\end{align}
Here, $\overline{\psi}_\mu:=\psi^T_\mu\sigma^2$, $\gamma^{\mu\nu\rho}:=\frac{-i}{4}\{[\gamma^\mu,\gamma^\nu],\gamma^\rho\}=e^\mu_ae^\nu_be^\rho_c\epsilon^{abc}$.
${\rm C}$ is the operator defined in eq.(\ref{C1}), and $D_{3/2,\nu}\psi$ is the covariant derivative on spin-$3/2$ fields:
\begin{align}
(D_{3/2,\nu}\psi)^\rho:=&\partial_\nu\psi^\rho+A_\nu\psi^\rho+\frac{1}{8}\omega_{\nu ab}[\sigma^a,\sigma^b]\psi^\rho+\Gamma_{\nu\rho\sigma}\psi^\sigma,\label{der}
\end{align}
where $\sigma^a$ ($a=1,2,3$) are the Pauli matrices eq.(\ref{Pauli}), $\Gamma^\mu{}_{\nu\rho}$ is the affine connection eq.(\ref{Affine}), and $A_\nu$ is a gauge field.
We use the spin connection eq.(\ref{1.5-order}) to satisfy $\delta S/\delta \omega_{\mu ab}=0$.
\footnote{
When we use the spin connection eq.(\ref{1.5-order}), (\ref{der}) cannot satisfy the linearity condition of the covariant derivative.
Here, we consider only the leading order of the coupling constant $\kappa$.
Then, eq.(\ref{der}) is linear.
}
The Rarita-Schwinger field $\psi_\mu$ is the gauge field of the following gauge transformation:
\begin{align}
\delta e^a{}_\mu=\frac{1}{2}(\overline{\epsilon}\gamma^a\psi_\mu-\overline{\psi}_\mu(\gamma^a)^\dag\epsilon),\qquad
\delta\psi_\mu=D_\mu\epsilon:=\partial_\mu\epsilon+\frac{1}{8}\omega_\mu{}^{ab}[\sigma_a,\sigma_b]\epsilon,
\label{local=SUSY}
\end{align}
where $\epsilon$ is a parameter of this transformation which is a spin-$1/2$ fermion.
We substitute gauge fixing functions \cite{Endo-Kimura}:
\begin{align}
G_1(\psi):=
i\gamma_\mu\psi^\mu-b,\qquad
G_2({\rm C}\psi):=i\gamma_\mu{\rm C}\psi^\mu-{\rm C}b
\end{align}
 into the path integral to define the partition function.
Here, $b$ is a spin-$1/2$ fermion.
Then, the partition function is given as follows \cite{Endo-Kimura, PVan}:
\begin{align}
Z[e]=&\int
{\cal D}\psi{\cal D}{\rm C}\psi
\left(\frac{1}{{\rm det}(i\slashed{D}_{1/2})}\right)^2
\int
{\cal D}F
{\cal D}{\rm C}F
\exp\left[-\frac{1}{2}\int d^3x\sqrt{g}\left\{
\overline{F}i\slashb{D}_{1/2}F
+\overline{{\rm C}F}i\slashb{D}_{1/2}{\rm C}F\right\}\right]
\notag \\
\times&
\exp\left[-\frac{1}{2}\int d^3x\: \sqrt{g}\:\overline{\psi}_\mu\gamma^{\mu\nu\rho}iD_{3/2,\nu}\psi_\rho
-\frac{1}{2}\int d^3x\: \sqrt{g}\:\overline{{\rm C}\psi}_\mu\gamma^{\mu\nu\rho}iD_{3/2,\nu}{\rm C}\psi_\rho
\right.
\notag \\
&
+\frac{1}{2}\sqrt{g}\overline{(\gamma_\rho\psi^\rho)} i\slashb{D_{1/2}}(\gamma_\mu\psi^\mu)
+\frac{1}{2}\sqrt{g}\overline{(\gamma_\rho{\rm C}\psi^\rho)} i\slashb{D}_{1/2}(\gamma_\mu{\rm C}\psi^\mu)
\notag \\
&\left.
-\frac{1}{2\kappa^2}\int d^3x\: \sqrt{g}\:e^{a\mu}e^{b\nu}R_{\mu\nu ab}\right].\label{pf-1}
\end{align}
Here, $i\slashed{D}_{1/2}$ is the spin-$1/2$ Dirac operator, $D_{3/2,\nu}$ is the covariant derivative eq.(\ref{der}), and $F$ is a spin-$1/2$ fermion.
We can generalize the symmetry eq.(\ref{local=SUSY}) to $F$ and the ghost term $\frac{1}{{\rm det}(i\slashed{D}_{1/2})}$ in (\ref{pf-1}) \cite{Nil-1,Nil-2,Nil-3,Kal,Hata-Kugo}.
In the following, we consider the leading contribution in $\kappa$ to the effective action (\ref{pf-1}).
Then, we can calculate anomalies of this theory by calculating anomalies of the first term $\left(\frac{1}{{\rm det}(i\slashed{D}_{1/2})}\right)^2$, the second term $\int
{\cal D}F
{\cal D}{\rm C}F
\exp\left[-\frac{1}{2}\int d^3x\sqrt{g}\left\{
\overline{F}i\slashb{D}_{1/2}F
+\overline{{\rm C}F}i\slashb{D}_{1/2}{\rm C}F\right\}\right]$, and other terms in the partition function eq.(\ref{pf-1}) independently.
In ordinary terminology, the total anomaly of the partition function eq.(\ref{pf-1}) is called the anomaly of the Rarita-Schwinger field.
But in this paper, the anomaly of the second and third lines of the partition function eq.(\ref{pf-1}) is denoted as the anomaly of the Rarita-Schwinger field.

We should define the partition function eq.(\ref{pf-1}) for a pseudo-real fermion.
Then, we need only one contribution from the pair $(F,CF)$ and its contribution is eq.(\ref{spin-1/2-pf}).
We also need only one contribution from the pair of the Kramers doubling of the ghost fields.
Then, the total contribution from the ghosts and $F$-field is the inverse of eq.(\ref{spin-1/2-pf}).

Let us finally consider anomalies of the Rarita-Schwinger field.
By using $\sigma_j^T=\sigma_j$ ($j=1,3$), $\sigma_2^T=-\sigma_2$, and $\gamma^\mu\gamma^\nu\gamma^\rho
=i\gamma^{\mu\nu\rho}
+g^{\mu\nu}\gamma^\rho
+g^{\nu\rho}\gamma^\mu
-g^{\mu\rho}\gamma^\nu$,
the effective Lagrangian of the Rarita-Schwinger field in eq.(\ref{pf-1}) is
\begin{align}
{\cal L}_\psi:=\frac{1}{2}\overline{\psi}_\mu({\cal R}\psi)^\mu
+\frac{1}{2}\overline{{\rm C}\psi}_\mu({\cal R}{\rm C}\psi)^\mu.
\end{align}
Here, the Rarita-Schwinger operator ${\cal R}$ is defined by using the spin-$3/2$ Dirac operator $i\slashed{D}_{3/2}$ as
\begin{align}
({\cal R}\psi)^\mu:=&
\gamma^{\mu\nu\rho}D_{3/2,\nu}\psi_\rho,
+\gamma^\mu i\slashed{D}_{1/2}(\gamma_\rho\psi^\rho)
=(i\slashed{D}_{3/2}\psi)^\mu-\gamma^\mu\gamma^\nu\psi^\rho(i\partial_\nu\gamma_\rho).
\label{RS-op}
\end{align}
From Lemma (B2) in Appendix B, we can show that the Rarita-Schwinger operator ${\cal R}$ and the spin-$3/2$ Dirac operator  $i\slashed{D}_{3/2}$ are both elliptic operators on $L^2(\mathbb{S}\otimes TM\otimes (P \times_G V))$.
The Rarita-Schwinger operator eq.(\ref{RS-op}) is self-adjoint:
\begin{align}
\left<{\cal R}\psi,\chi\right>=&
\left<\psi,{\cal R}\chi\right>.
\label{SA-RS}
\end{align}
All eigenmodes ${\cal R}\psi_j=\lambda_j\psi_j$ satisfy
\begin{align}
\left<\psi_j,\psi_k\right>=\delta_{jk},\qquad
\left<\psi_j,{\rm C}\psi_k\right>=0.\label{RS-base1}
\end{align}
We choose a complex basis of the Rarita-Schwinger field as $\psi_j$ and ${\rm C}\psi_j$.
After adding the Pauli-Villars regulator, the partition function of the Rarita-Schwigner field becomes
\begin{align}
Z_c^{3/2}=\prod_j\frac{\lambda_j}{\lambda_j+iM}=\exp(i\pi\eta({\cal R}(A,g))).
\end{align}
Here, we denote $\eta({\cal R}(A,g))$ the (reduced) $\eta$-invariant of the Rarita-Schwinger operator ${\cal R}(A,g)$ with a gauge field $A$ and a metric $g$ defined in eq.(\ref{eta=def}).
Since the Rarita-Schwinger field is pseudo-real, we define the effective action by only one of the contributions from the pair $(\psi_j,{\rm C}\psi_j)$:
\begin{align}
Z^{3/2}=\exp(i\pi\eta({\cal R}(A,g))/2).\label{pf-gravitino}
\end{align}
Since the Pauli-Villars regulator does not break any symmetries, there are no perturbative anomalies.
We will discuss global anomalies.

Let us consider the difference between $\eta({\cal R}(A,g))$ and $\eta({\cal R}(A^\pi,g^\pi))$, where $A^\pi$ and $g^\pi$ are a gauge field and a metric on $X$ after a gauge-diffeomorphism transformation $\pi$.
Since the Rairta-Schwinger operator is an elliptic self-adjoint operator (See Lemma (B2) of Appendix B), we can use eq.(\ref{dif-eta-1}) and obtain
\begin{align}
\eta({\cal R}(A^\pi,g^\pi))-\eta({\cal R}(A,g))={\rm SF}(f_{R\to R_\pi}).
\label{GA-flow}
\end{align}
Here, we denote $f_{R\to R_\pi}$ a path on the set of the elliptic operators that connect ${\cal R}(A,g)$ and ${\cal R}(A^\pi,g^\pi)$:
\begin{align}
f_{R\to R_\pi}:[0,1]\to{\cal CF}^{sa}(L^2(\mathbb{S}\otimes TM\otimes (P \times_G V)));\qquad t\mapsto {\cal R}(A_t,g_t),\label{R-path}
\end{align}
where ${\cal R}(A_t,g_t)$ is the Rarita-Schwinger operator eq.(\ref{RS-op}) with a gauge connection $A_t$ and a metric $g_t$ defined in eq.(\ref{flow-path-2}), and the set ${\cal CF}^{sa}(L^2(\mathbb{S}\otimes TM\otimes (P \times_G V)))$ is defined in eq.(\ref{fredholm}).
Let us check that $f_{R\to R_\pi}$ is continuous with the gap phase.
Consider an open neighborhood $U_\epsilon({\cal R}(A_t,g_t))\subset {\cal CF}^{sa}(L^2(\mathbb{S}\otimes TM\otimes (P \times_G V)))$ of a point ${\cal R}(A_t,g_t)$.
Here, $U_\epsilon({\cal R}(A_t,g_t))$ is defined in eq.(\ref{OS-gap}).
Then, we need to find a constant $\delta>0$ such that ${\cal R}(A_{t'},g_{t'})\in U_\epsilon({\cal R}(A_t,g_t))$ for any $t'$ with $|t-t'|<\delta$.
We find
\begin{align}
d({\cal R}(A_t,g_t),\:{\cal R}(A_{t'},g_{t'}))
={\rm sup}_{u\neq 0}
\frac{2\left|\left|\left\{({\cal R}(A_t,g_t)+i)^{-1}-({\cal R}(A_{t'},g_{t'})+i)^{-1}\right\}u\right|\right|}{||u||}.\label{cont-2}
\end{align}
Let us denote $({\cal R}(A_t,g_t)+i)^{-1}u=v_1$ and $({\cal R}(A_{t'},g_{t'})+i)^{-1}u=v_1+v_2$.
Then, we obtain
\begin{align}
\left\{{\cal R}(A_t,g_t)-{\cal R}(A_{t'},g_{t'})\right\}v_1=({\cal R}(A_{t'},g_{t'})+i)v_2.\label{cont-1}
\end{align}
On the other hand, by the definition (\ref{RS-op}), we find
\begin{align}
{\cal R}(A_{t+\delta},g_{t+\delta })={\cal R}(A_t,g_t)+\delta \Delta{\cal R}(\delta,A_t,g_t),
\end{align}
where $\Delta{\cal R}(\delta,A_t,g_t)$ is a bounded operator which satisfies $\Delta{\cal R}(\delta,A_t,g_t)\to {\rm const}$ ($\delta \to 0$).
Therefore, by using (\ref{cont-1}), we find $v_2={\cal O}(\delta)$.
Since the right hand side of eq.(\ref{cont-2}) is ${\rm sup}_{u\neq 0}\frac{2||v_2||}{||u||}$, we finally obtain that 
\begin{align}
d({\cal R}(A_t,g_t),\:{\cal R}(A_{t'},g_{t'}))
={\cal O}(|t-t'|).\label{cont-3}
\end{align}
This is the proof of continuity of the map $f_{R\to R_\pi}$.
It is difficult to identify the partition function (\ref{pf-gravitino}) as the path integral of some anomaly-free bulk system because the Rarita-Schwinger operator is not divisible to the two parts, one simply includes the differential of the cylinder direction, and the other part does not depend on the cylinder direction on the mapping torus.

As we mentioned in Appendix B, the spin-$3/2$ Dirac operator is a self-adjoint elliptic operator.
We will show that the spectral flow of the spin-$3/2$ Dirac operator is equivalent to that of the Rarita-Schwinger operator \cite{Freed-K} by using some results in Appendix B.
Consider a path on elliptic operators that connect spin-$3/2$ Dirac operators $i\slashed{D}_{3/2}(A,g)$ and $i\slashed{D}_{3/2}(A^\pi,g^\pi)$:
\begin{align}
f_{D^{3/2}\to D^{3/2}_\pi}:[0,1]\to{\cal CF}^{sa}(L^2(\mathbb{S}\otimes TM\otimes (P \times_G V))),\qquad t\mapsto i\slashed{D}_{3/2}(A_t,g_t).\label{3/2-path}
\end{align}
Here, $i\slashed{D}_{3/2}(A_t,g_t)$ is the spin-$3/2$ Dirac operator with a gauge connection $A_t$ and a metric $g_t$ defined in eq.(\ref{flow-path-2}).
We can show that the path eq.(\ref{3/2-path}) is continuous in the same way as the proof of continuity for the path $f_{R\to R_\pi}$ in eq.(\ref{R-path}).
We also consider two paths $f_1,f_2:\:[0,1]\to{\cal CF}^{sa}(L^2(\mathbb{S}\otimes TM\otimes (P \times_G V)))$:
\begin{align}
f_1(s)=& si\slashed{D}_{3/2}(A,g)+(1-s){\cal R}(A,g)
=-i\slashed{D}_{3/2}(A,g)+(1-s)\gamma^\mu\gamma^\nu(i\partial_\nu\gamma_\rho)_{(g)},
\notag \\
f_2(s)=& s{\cal R}(A^\pi,g^\pi)+(1-s)i\slashed{D}_{3/2}(A^\pi,g^\pi)
=-i\slashed{D}_{3/2}(A^\pi,g^\pi)+(1-s)\gamma^\mu\gamma^\nu(i\partial_\nu\gamma_\rho)_{(g^\pi)}.
\label{path-2}
\end{align}
Here, we use eq.(\ref{RS-op}), and we denote $(i\partial_\nu\gamma_\rho)_{(g)}$ the term $i\partial_\nu\gamma_\rho$ with a metric $g_{\mu\nu}$.
We can confirm that $f_1(s)$ and $f_2(u)$ are paths on the set of self-adjoint Fredholm operators in the same way as the proof of Lemma (B2) in Appendix B
(The first term in the second line of eq.(\ref{path-2}) is an order-$1 $ elliptic operator, and the second term is pseudo-differential of order zero).
We can also check that two paths $f_1$ and $f_2$ are continuous in the same way as the continuity of the path $f_{R\to R_\pi}$ in eq.(\ref{R-path}).
Combining eq.(\ref{R-path}), eq.(\ref{3/2-path}), and eq.(\ref{path-2}), we obtain two continuous paths:
\begin{align}
F(s):=&\left\{
\begin{array}{ccc}
f_1(3s),&\qquad &s\in[0,1/3],\\
f_{D^{3/2}\to D^{3/2}_\pi}(3s-1),&&s\in[1/3,2/3],\\
f_2(3s-2),&&s\in[2/3,1],
\end{array}
\right.
\notag \\
G(s):=&\left\{
\begin{array}{ccc}
{\cal R}(A,g),&\qquad &s\in[0,1/3],\\
f_{R\to R_\pi}(3s-1),&&s\in[1/3,2/3],\\
{\cal R}(A^\pi,g^\pi),&&s\in[2/3,1].
\end{array}
\right.
\label{hom-fred}
\end{align}
We will show that these two maps $F(s)$ and $G(s)$ are homotopy equivalent.
We consider a continuous map:
\begin{align}
\Phi:\:[0,1]\times[0,1]\to{\cal CF}^{sa}(L^2(\mathbb{S}\otimes TM\otimes (P \times_G V)))
;\qquad (s,u)\mapsto (1-u)G(s)+uF(s).
\end{align}
$\Phi$ is valued in self-adjoint operators.
We will confirm this map is valued in the set of Fredholm operators.
In the case $0\leq s\leq 1/3$, 
\begin{align}
\Phi(s,u)=&-i\slashed{D}_{3/2}(A,g)+(1-3su)\gamma^\mu\gamma^\nu(i\partial_\nu\gamma_\rho)_{(g)}.
\end{align}
Since the first term is an order-$1 $ elliptic operator, and the second term is a pseudo-differential of order zero, $\Phi(s,u)$ is elliptic if $0\leq s\leq 1/3$ (See Lemma (B1) of Appendix B).
In the same way, $\Phi(s,u)$ is elliptic for all $s\in[0,1]$.
Therefore, $\Phi(s,u)$ is valued in the set of self-adjoint Fredholm operators.
We can also confirm that the map $\Phi(s,u)$ is continuous, and thus $\Phi(s,u)$ is the homotopy map that connects $G(s)$ and $F(s)$.
Thus, the spectral flow of $G(s)$ and $F(s)$ are equivalent (See Lemma (B3) in Appendix B).
Using the definition of the spectral flow (\ref{sflow-2}), we find
\begin{align}
{\rm SF}(f_1)+{\rm SF}\left(f_{D^{3/2}\to D^{3/2}_\pi}\right)+{\rm SF}(f_2)={\rm SF}\left(f_{R\to R_\pi}\right).
\label{flow-sum}
\end{align}
We will now show that ${\rm SF}(f_1)=-{\rm SF}(f_2)$.
For convenience, we introduce
\begin{align}
f_3(s):=f_2(1-s),\qquad s\in[0,1].
\end{align}
Then, the spectral flow satisfies ${\rm SF}(f_3)=-{\rm SF}(f_2)$.
We can choose a partition $\{0=t_0<t_1<\cdots<t_n=1\}$ of the interval $[0,1]$ and real numbers $\epsilon_j>0$ that satisfies the condition eq.(\ref{sflow-1}) for $f_1$ and $f_2$.
We obtain maps $E_j$ and $\tilde{E}_j$ from $[t_{j-1},t_j]$ to the set of bounded linear operators as eq.(\ref{sflow-1}), which correspond to above two paths $f_1$ and $f_3$:
\begin{align}
E_j:&\:[t_{j-1},t_j]\to\{{\rm bounded}\:{\rm linear}\:{\rm operators}\};\qquad t\mapsto\frac{1}{2\pi i}\int_{\Gamma_j}(\lambda-f_1(t))^{-1}d\lambda,
\notag \\
\tilde{E}_j:&\:[t_{j-1},t_j]\to\{{\rm bounded}\:{\rm linear}\:{\rm operators}\};\qquad t\mapsto\frac{1}{2\pi i}\int_{\Gamma_j}(\lambda-f_3(t))^{-1}d\lambda.
\end{align}
Here, we  denote $\Gamma_j$ the circle of radius $(\epsilon_j-\epsilon_{j-1})$ and center $(\epsilon_j+\epsilon_{j-1})/2$.
Therefore, the spectral flows of the two paths $f_1$ and $f_3$ are equivalent:
\begin{align}
{\rm SF}(f_1):=&\sum_{j=1}^n\left\{
{\rm dim}\left(E^\geq_j(t_j)\right)-{\rm dim}\left(E^\geq_j(t_{j-1})\right)\right\}
\notag \\
=&\sum_{j=1}^n\left\{
{\rm dim}\left(\tilde{E}^\geq_j(t_j)\right)-{\rm dim}\left(\tilde{E}^\geq_j(t_{j-1})\right)\right\}
={\rm SF}(f_3)=-{\rm SF}(f_2).
\label{flow-zero}
\end{align}
Combining eq.(\ref{flow-sum}) and eq.(\ref{flow-zero}), the spectral flow of spin-$3/2$ Dirac operator and the Rarita-Schwinger operator are equivalent:
\begin{align}
{\rm SF}\left(f_{D^{3/2}\to D^{3/2}_\pi}\right)={\rm SF}\left(f_{R\to R_\pi}\right).
\label{flow-equal}
\end{align}
Since the spin-$3/2$ Dirac operator is elliptic, eq.(\ref{dif-eta-1}) holds for the spin-$3/2$ Dirac operator.
Therefore, eq.(\ref{flow-equal}) is similar to \cite{FH-19}.
By the Theorem 7.4 of \cite{APS3}, we obtain ${\rm SF}\left(f_{D^{3/2}\to D^{3/2}_\pi}\right)={\cal I}_{3/2}(\pi)$, where ${\cal I}_{3/2}(\pi)$ is the index of the spin-$3/2$ Dirac operator on the mapping torus of $\pi$.
Combining this fact and eq.(\ref{pf-gravitino}), eq.(\ref{GA-flow}), and eq.(\ref{flow-equal}), we finally obtain that the global anomaly for a gauge-diffeomorphism transformation $\pi$ of the Rarita-Schwinger field  is given by $(-1)^{{\cal I}_{3/2}(\pi)/2}$.

We will finally see that global anomalies of the Rarita-Schwinger theory are captured by the anomaly inflow.
We consider a Euclidian $4$-dimensional spin manifold $Y$ whose boundary is the original manifold $\partial Y=X$ and assume that $Y$ is a cylinder near the boundary.
We introduce the coordinates $(x_0,x_1,x_2,x_3)$.
Here, $x_0\in[0,1]$ is the cylinder direction.
Let us consider a Lagrangian ${\cal L}={\cal L}_{3/2}+{\cal L}_1+{\cal L}_2$.
Here, ${\cal L}_{3/2}$ is the Lagrangian of a massive spin-$3/2$ Dirac field $\Psi^\mu(x)$ with fermionic statistics and ${\cal L}_j$ ($j=1,2$) are two Lagrangians of a massive spin-$1/2$ Dirac field $\chi_1,\chi_2$ with bosonic statistics:
\footnote{
If we consider only one spin-$1/2$ Dirac boson, we can obtain the anomaly inflow of the Rarita-Schwinger operator ${\cal R}$.
}
\begin{align}
{\cal L}_{3/2}=&\overline{\Psi}_\mu( i\slashed{D}_{3/2}+iM)\Psi^\mu
+\overline{{\rm C}\Psi}_\mu( i\slashed{D}_{3/2}+iM){\rm C}\Psi^\mu,
\notag \\
{\cal L}_j=&\overline{\chi}_j( i\slashed{D}_{1/2}+iM)\chi_j
+\overline{{\rm C}\chi}_j( i\slashed{D}_{1/2}+iM){\rm C}\chi_j,\qquad j=1,2.\label{M-grav}
\end{align}
Here, $\overline{\Psi}_\mu:=\Psi_\mu^T\Gamma_2$, $\overline{\chi}_j:=\chi_j^T\Gamma_2$,
$i\slashed{D}_{3/2}$ is the spin-$3/2$ Dirac operator, $\Gamma^\mu$ is the gamma matrices defined in eq.(\ref{4d-gamma}), $M>0$ the mass, and ${\rm C}$ is defined in eq.(\ref{C-4d}).
We assume that those fields $\Psi_\mu$, $\chi_j$ ($j=1,2$) are defined by a common spin structure on $Y$ and the spin connection in (\ref{M-grav}) becomes eq.(\ref{1.5-order}) with ignoring the torsion part near the boundary.
Since the metric near the boundary does not depend on $x_0$, the components of the Affine connection that include the $x_0$ direction are zero near the boundary ($\Gamma^0{}_{\mu\nu}=\Gamma^\mu{}_{0\nu}=0$).
Thus, near the boundary, the Lagrangian ${\cal L}_{3/2}$ becomes
\begin{align}
{\cal L}_{3/2}
=
\overline{\Psi}_\mu
i\left(
\Gamma^0\frac{\partial}{\partial x_0}
+\sum_{\nu\neq 0}\Gamma^\nu D_{3/2,\nu}
+ M\right)\Psi^\mu
+
\overline{{\rm C}\Psi}_\mu
i\left(
\Gamma^0\frac{\partial}{\partial x_0}
+\sum_{\nu\neq 0}\Gamma^\nu D_{3/2,\nu}
+ M\right){\rm C}\Psi^\mu,
\end{align}
where $D_{3/2,\nu}$ is the covariant derivative eq.(\ref{der}) in the $\nu$-direction.
Let us choose the boundary condition $L$ as follows:
\begin{align}
L:\:(1-\Gamma^5)\Psi_\mu(x)|_{x^0=0}=0,\qquad
(1-\Gamma^5)\chi_j(x)|_{x^0=0}=0,\qquad j=1,2.\label{inflow-bc}
\end{align}
In the same way as \cite{WY}, we find that the original massless spin-$3/2$ Dirac fermion, an additional massless spin-$1/2$ Dirac fermion, and two massless spin-$1/2$ Dirac bosons appear as localized modes on the boundary.

We will now determine the phase of the path integral of the bulk system on a $4$-dimensional manifold $Y$.
Since the spin-$3/2$ and spin-$1/2$ Dirac operators are not self-adjoint on $X$ with the boundary condition eq.(\ref{inflow-bc}), we cannot choose the eigenmodes of these Dirac operators as a complex basis of spin-$3/2$ and spin-$1/2$ fields.
But in the same way as \cite{WY}, we find 
\begin{align}
Z_c^{3/2}(Y|L)=\left<L|\Omega\right>\left<\Omega|Y\right>.\label{3/2-bulk-1}
\end{align}
Here, $Z_c^{3/2}(Y|L)$ is the path integral around $Y$ with the boundary condition eq.(\ref{inflow-bc}).
$\bra{L}$ is the state that corresponds to the boundary condition eq.(\ref{inflow-bc}) and $\left<\Omega|Y\right>$ is the path integral over $Y$ with a boundary condition that corresponds to the vacuum $\ket{\Omega}$ \cite{KY,WY}.
In the case there are no zero modes on the boundary, we can consider the APS boundary condition \cite{APS1}. 
Since $\Gamma^0{}_{\mu\nu}=\Gamma^\mu{}_{0\nu}=0$, the spin-$3/2$ Dirac operator can be divided into two terms in the same as a spin-$1/2$ fermion near the boundary.
Therefore, we can use section $2$ of \cite{KY}, and we find that the state $\ket{{\rm APS}}$ which corresponds to the APS boundary condition is equivalent to the vacuum state of the massless theory $M=0$ in eq.(\ref{M-grav}).
Furthermore, if there are no zero modes in the boundary theory, there are unique vacuum on the boundary.
Therefore, $\ket{Y}$ is multiple of $\ket{\Omega}$, and then we find that $\ket{Y}$ is multiple of $\ket{\Omega}$ (See section 2 of \cite{WY}).
Eq.(\ref{3/2-bulk-1}) becomes
\begin{align}
Z_c^{3/2}(Y|L)
=
\frac{\left<L|\Omega\right>\left<\Omega|{\rm APS}\right>}{|\left<{\rm APS}|\Omega\right>|^2}
\left<{\rm APS}|Y\right>.\label{DF-KY}
\end{align}
The first term does not have any phase \cite{WY}.
Since eq.(\ref{DF-KY}) is diverse, we should introduce the Pauli-Villas regulator fields whose masses are $-M$.
Since the boundary condition eq.(\ref{inflow-bc}) and APS boundary condition do not depend on the mass of the bulk system, states $\ket{L}$ and $\ket{{\rm APS}}$ do not depend on mass.
Therefore, there is no phase in the first term after adding the Pauli-Villars regulator fields.
Now we consider the phase of the second term in eq.(\ref{DF-KY}).
The spin-$3/2$ and spin-$1/2$ Dirac operators are self-adjoint with the APS boundary condition \cite{APS1,KY}.
Then, we can calculate the path integral $\left<{\rm APS}|Y\right>$ in the same way as the calculation of a partition function on the boundary theory, and we find that the phase of the second term in eq.(\ref{DF-KY}) is given as $\exp\left(2i\pi\eta(i\slashed{D}_{3/2}(Y|{\rm APS}))\right)\cdot\exp\left(-2i\pi\eta(i\slashed{D}_{1/2}(Y|{\rm APS}))\right)$, where $i\slashed{D}_{3/2}(Y|{\rm APS})$ and $i\slashed{D}_{1/2}(Y|{\rm APS})$ are the spin-$3/2$ Dirac operator and the spin-$1/2$ Dirac operator on $Y$ with the APS boundary condition.
Thus, if we choose one of the contributions from each pair of the Kramers doubling by the operator ${\rm C}$ in the path integral (\ref{DF-KY}), we obtain the path integral over the pseudo-real fermion on $Y$ \cite{WY}:
\begin{align}
Z^{3/2}(Y|L)=|Z^{3/2}(Y|L)|\exp\left(i\pi\eta(i\slashed{D}_{3/2}(Y|{\rm APS}))\right)
\cdot\exp\left(-i\pi\eta(i\slashed{D}_{1/2}(Y|{\rm APS}))\right).\label{inflow-phase}
\end{align}
If there exist zero modes of the spin-$3/2$ Dirac operator or the spin-$1/2$ Dirac operator on $Y$, we should insert all zero modes in the path integral and we can obtain the same phase of eq.(\ref{inflow-phase}) \cite{WY}.

Now we consider a gauge-diffeomorphism transformation $\pi$ on the boundary $X=\partial Y$ of the bulk system.
The bulk is changed as $Y\to Y+M_\pi$ by $\pi$ \cite{KY}, where $M_\pi$ is the corresponding mapping torus, and $Y+M_\pi$ is the connected sum of $Y$ and $M_\pi$ with the common boundary $X$.
Then, the path integral (\ref{inflow-phase}) is changed by a gauge-diffeomorphism transformation $\pi$ on the boundary as
\begin{align}
Z^{3/2}(Y|L)\to& Z^{3/2}(Y+M_\pi|L)
\notag \\
=&|Z^{3/2}(Y|L)|\exp\left(i\pi\eta(i\slashed{D}_{3/2}(Y+M_\pi|{\rm APS}))\right)
\cdot\exp\left(-i\pi\eta(i\slashed{D}_{1/2}(Y|{\rm APS}))\right)
\notag \\
=&\exp\left(i\pi\eta(i\slashed{D}_{3/2}(M_\pi))\right)
\cdot\exp\left(-i\pi\eta(i\slashed{D}_{1/2}(M_\pi))\right)
Z^{3/2}(Y|L).\label{AP-RS}
\end{align}
In the third line, we use the Dai-Freed theorem \cite{Dai-Freed}.
Since all non-zero eigenvalues of the spin-$3/2$ and spin-$1/2$ Dirac operators on the mapping torus $M_\pi$ make quartets $(\lambda,\lambda,-\lambda,-\lambda)$ by the existence of the operators ${\rm C}$ and $\Gamma^5$, we find $\exp\left(i\pi\eta(i\slashed{D}_{3/2}(M_\pi))\right)=(-1)^{N_0/2}$ and $\exp\left(i\pi\eta(i\slashed{D}_{1/2}(M_\pi))\right)=(-1)^{n_0/2}$, where $N_0$ and $n_0$ are the number of zero modes pf the spin-$3/2$ and spin-$1/2$ Dirac operators on the mapping torus $M_\pi$.
Furthermore, since $[{\rm C},\Gamma^5]=0$, we find that the number of zero modes with positive chirality and the number of zero modes with negative chirality are both even for the spin-$3/2$ and spin-$1/2$ Dirac operators.
Therefore, we find that this change can be written as $\exp\left(i\pi\eta(i\slashed{D}_{3/2}(M_\pi))\right)\cdot\exp\left(-i\pi\eta(i\slashed{D}_{1/2}(M_\pi))\right)=(-1)^{({\cal I}_{3/2}(\pi)+{\cal I}_{1/2}(\pi))/2}$, which means that this change of the bulk path integral is equivalent to the global anomaly of the transformation $\pi$ in the original $3$-dimensional system.
Therefore, we can identify the path integral over the bulk system with a boundary condition eq.(\ref{inflow-bc}) and the partition function of the original $3$-dimensional theory (\ref{pf-1}), when we consider global anomalies of the $3$-dimensional original theory.
We can also obtain the anomaly inflow in the case the gauge group is $U(1)$.

\subsection{${\rm Pin}^+$ Unorientable Manifold}
We now consider the Rarita-Schwinger theory on an unorientable ${\rm Pin}^+$ manifold $X$, which means $w_1(X)\neq 0$ and $w_2(X)=0$.
As explained in section $2$, we can write $X=\hat{X}/\tau$, where $\hat{X}$ is an orientable double cover of $X$ which is a spin manifold, and $\tau:\hat{X}\to \hat{X}$ is an orientation reversing that satisfies $\tau^2=1$.
It is enough to choose $\tau$ as $\tau:x^\mu\to x^\mu-2w^\mu(x\cdot w)$, where $w=w^\mu\partial_\mu$ is a tangent vector \cite{EW}.
$\tau$ acts on spinors via the frame field.
A spinor $\psi^\mu$ on $\hat{X}$ is transformed by a transformation $P^\mu{}_\nu\in O(3)$ on the frame bundle as 
\begin{align}
\psi^\mu(x)\to \hat{S}(P)[\psi]^\mu(x):=
\left\{
\begin{array}{c}
{\cal S}(P)P^\mu{}_\nu\psi^\nu(Px),
\\
{\rm det}\left({\cal S}(P)\right)\cdot{\cal S}(P)P^\mu{}_\nu\psi^\nu(Px).
\end{array}
\right.\label{spin3/2-un}
\end{align}
Here, we use the transformation rule eq.(\ref{UO-3}) and use the notation eq.(\ref{UO-4}).
A field $\psi^\mu$ that transformed by the first rule of eq.(\ref{spin3/2-un}) is an element of the Hilbert space $L^2({\cal P}_+\otimes T\hat{X}\otimes (P \times_G V))$, and a field $\psi^\mu$ that transformed by the second rule of eq.(\ref{spin3/2-un}) is an element of the Hilbert space $L^2({\cal P}_-\otimes T\hat{X}\otimes (P \times_G V))$, where ${\cal P}_\pm$ are defined in below of eq.(\ref{UO-3}), and we denote $T\hat{X}$ the tangent bundle on $\hat{X}$.
$P \times_G V\to\hat{X}$ is a $G$-bundle with a real representation of a Lie group $G$ on a linear space $V$.
Since the inner product eq.(\ref{inner-RS}) is invariant under $\tau$, we can define Lebesgue inner product $\left<\psi_\pm,\psi_\pm\right>$ for $\psi_\pm^\mu\in L^2({\cal P}_\pm\otimes T\hat{X}\otimes (P \times_G V))$.
Since $\left<\tau\psi,\chi\right>=\left<\psi,\tau\chi\right>$, we find 
\begin{align}
\left<\psi_+,\chi_-\right>=0,\qquad\forall \psi_+\in L^2({\cal P}_+\otimes T\hat{X}\otimes (P \times_G V)),\qquad
\forall \chi_-\in L^2({\cal P}_-\otimes T\hat{X}\otimes (P \times_G V)).
\end{align}
We first construct Rarita-Schwinger Lagrangian that includes only $L^2({\cal P}_+\otimes T\hat{X}\otimes (P \times_G V))$.
We first note that the operator $\psi^\mu\to \gamma^{\mu\nu\rho}(D_\nu\psi)_\rho$ is anti-commute with the reflection $\tau$.
This is because the $\gamma^{\mu\nu\rho}$ is anti-symmetric for $\mu,\nu,\rho$, and thus we confirm this fact in the same way as the Dirac operator.
Furthermore, $\psi_\pm^T\sigma^2\in L^2({\cal P}_\mp\otimes T\hat{X}\otimes (P \times_G V))$, where $\psi_\pm\in L^2({\cal P}_\pm\otimes T\hat{X}\otimes (P \times_G V))$ \cite{EW}.
Therefore, we can construct the Rarita-Schwinger Lagrangian in Euclidian signature only the contribution from $L^2({\cal P}_+\otimes T\hat{X}\otimes (P \times_G V))$:
\begin{align}
S^E_{\rm SUGRA}=&S^E_2+S^E_{3/2},
\notag \\
S^E_2=&-\frac{1}{4\kappa^2}\int d^3x\:  \sqrt{g}\:e^{a\mu}e^{b\nu}R_{\mu\nu ab},
\notag \\
S^E_{3/2}=&\frac{1}{2}\int d^3x\:  \sqrt{g}\:\overline{\psi_+}_{\mu}\gamma^{\mu\nu\rho}(D_{3/2,\nu}\psi_+)_\rho.
\label{SGU=Action}
\end{align}
Here, $\overline{\psi}_{+\mu}:=\psi^T_{+\mu}\sigma^2$, $D_{3/2,\nu}\psi_+$ is the covariant derivative of the Rarita-Schwinger field eq.(\ref{der}), and $\psi_+\in L^2({\cal P}_+\otimes T\hat{X}\otimes (P \times_G V))$.
We can confirm this action is invariant by $\tau$.
In the following, we will use the spin connection eq.(\ref{1.5-order}).
We should also confirm that all symmetries are closed on $L^2({\cal P}_+\otimes T\hat{X}\otimes (P \times_G V))$ or not.
It is obvious that any gauge symmetries and Lorentz transformations are closed on the space $L^2({\cal P}_+\otimes T\hat{X}\otimes (P \times_G V))$.
The transformation eq.(\ref{local=SUSY}) is also closed on $L^2({\cal P}_+\otimes T\hat{X}\otimes (P \times_G V))$ if we assume the transformation parameter as $\epsilon\in L^2({\cal P}_+)$.

Let us consider the partition function of the Rarita-Schwinger theory.
To define the partition function, we should first insert the gauge-fixing function:
\begin{align}
G_1(\psi_+):=i\gamma_\mu\psi_+^\mu-b,\qquad
 \psi_+\in L^2({\cal P}_+\otimes T\hat{X}\otimes (P \times_G V))\label{GF-un3/2}
\end{align}
into the path integral.
Since $\gamma_\mu\psi_+^\mu\in L^2({\cal P}_+\otimes (P \times_G V))$, we choose $b\in L^2({\cal P}_+\otimes (P \times_G V))$.
 We can confirm that this gauge-fixing condition is invariant under $\tau$.
Then, the partition function of eq.(\ref{SGU=Action}) is given as 
\begin{align}
Z[e]=&\int
{\cal D}\psi_+
\frac{1}{{\rm det}(i\slashed{D}_{1/2})}
\int
{\cal D}F_+
\exp\left[-\frac{1}{2}\int d^3x\sqrt{g}\overline{F_+}i\slashed{D}_{1/2}F_+
\right]
\notag \\
\times&
\exp\left[-\frac{1}{2}\int d^3x\sqrt{g}
\overline{\psi}_{+\mu}\gamma^{\mu\nu\rho}iD_{3/2,\nu}\psi_{+\rho}
+\frac{1}{2}\int d^3x\sqrt{g}
\overline{(\gamma_\rho\psi_+^\rho)} i\slashed{D}_{1/2}(\gamma_\mu\psi_+^\mu)
\right.
\notag \\
&\left.-\frac{1}{2\kappa^2}\int d^3x\: \sqrt{g}\:e^{a\mu}e^{b\nu}R_{\mu\nu ab}\right].\label{pfU-1}
\end{align}
Here, $\psi^\mu_+\in L^2({\cal P}_+\otimes T\hat{X}\otimes (P \times_G V))$, and $F_+\in L^2({\cal P}_+\otimes  (P \times_G V))$.
We will consider the leading order of the effective action eq.(\ref{pfU-1}) for the coupling constant $\kappa$.
Then, we will ignore the torsion part of the spin connection eq.(\ref{1.5-order}).
The product of the term $({\rm det}(i\slashed{D}_{1/2}))^{-1}$ and the path integral $\int
{\cal D}F_+
\exp\left[-\frac{1}{2}\int d^3x\sqrt{g}\overline{F_+}i\slashed{D}_{1/2}F_+
\right]$ in eq.(\ref{pfU-1}) is totally the inverse of eq.(\ref{pin+pf}).
Therefore, their contribution to the global anomalies are captured by the spectral flow of the spin-$1/2$ Dirac operator as explained in section $2.2$.

Let us consider the second line of the effective Lagrangian eq.(\ref{pfU-1}):
\begin{align}
{\cal L}_+=\overline{\psi}_{+\mu}({\cal R}\psi_+)^\mu,\label{L-RS=Pin}
\end{align}
where ${\cal R}$ is defined in eq.(\ref{RS-op}).
We can confirm that this operator ${\cal R}$ commutes with ${\rm C}$ defined in eq.(\ref{C1}).
We now choose a basis of the space $L^2({\cal P}_+\otimes T\hat{X}\otimes (P \times_G V))$.
Since $\{\tau,{\cal R}\}=0$, ${\cal R}$ maps from $L^2({\cal P}_+\otimes T\hat{X}\otimes (P \times_G V))$ to $L^2({\cal P}_-\otimes T\hat{X}\otimes (P \times_G V))$, we cannot choose the basis of the space $L^2({\cal P}_+\otimes T\hat{X}\otimes (P \times_G V))$ as the eigenmodes of ${\cal R}$.
However, since ${\cal R}$ is close and self-adjont on the space $L^2(({\cal P}_+\oplus{\cal P}_-)\otimes T\hat{X}\otimes (P \times_G V))$, we choose the basis of the space $L^2(({\cal P}_+\oplus{\cal P}_-)\otimes T\hat{X}\otimes (P \times_G V))$ as all eigenmodes of the operator ${\cal R}\psi_j=\lambda_j\psi_j$, $\lambda_j\in\mathbb{R}$.
We divide each eigenmode as
\begin{align}
\psi_j=\psi_{j,+}+\psi_{j,-}\qquad\psi_{j,\pm}\in L^2({\cal P}_\pm\otimes T\hat{X}\otimes (P \times_G V)).
\end{align}
 Then, the eigenvalue of $\tau\psi_j=\psi_{j,+}-\psi_{j,-}$ is $-\lambda_j$.
 Therefore, we choose the basis of the space as the set $\{\psi_{j,+},\psi_{j,-}\}_{\lambda_j>0}$ and zero modes.
 By using $\left<\psi_j,\tau\psi_k\right>=0$ where the eigenvalue of $\psi_j$ and $\psi_k$ are positive, and use $\left<\psi_j,\psi_k\right>=2\delta_{jk}$, we find
 \begin{align}
 \left<\psi_{j,+},\psi_{k,+}\right>=\left<\psi_{j,-},\psi_{k,-}\right>=\delta_{jk},\qquad
 \left<\psi_{j,+},\psi_{k,-}\right>=0,\qquad \lambda_j,\lambda_k>0.
 \end{align}
Since ${\cal R}$ and $\tau$ are commute on the subspace of zero modes, we choose the complex basis of the space $L^2({\cal P}_\pm\otimes T\hat{X}\otimes (P \times_G V))$ as the set $\{\psi_{j,\pm}\}_{\lambda_j>0}\oplus \{\phi_{A,\pm}\}$, where $\phi_{A,\pm}$ are zero modes with $\tau\phi_{A,\pm}=\pm\phi_{A,\pm}$.

Now we will consider the partition function of the Lagrangian eq.(\ref{L-RS=Pin}).
In the following, we will regard the partition function of the Lagrangian eq.(\ref{L-RS=Pin}) to be the partition function of the Rarita-Schwinger field.
\footnote{
In ordinary terminology, the partition function of the total effective Lagrangian eq.(\ref{pfU-1}) is called the partition function of the Rarita-Schwinger field.
}
We assume that there are no zero modes.
(If there exist zero modes, we insert all zero modes into the path integral.)
We expand the Rarita-Schwinger field in $L^2({\cal P}_+\otimes T\hat{X}\otimes (P \times_G V))$ as
\begin{align}
\psi_+=&\sum_ja_j\psi_{j,+},\qquad {\cal R}\psi_{j,+}=\lambda_j\psi_{j,-},\qquad \lambda_j>0,\qquad a_j\in\mathbb{C}.
\end{align}
The partition function is given as
\begin{align}
Z_+=&
\int da_{j,R}da_{j,I}
\exp\left(
i\sum_{j,k;\lambda_j,\lambda_k>0}a_{j,R}(\lambda_j B_{jk}-\lambda_kB_{kj})a_{k,I}
\right)
\end{align}
Here, $a_j=a_{j,R}+ia_{j,I}$ and $B_{jk}:=\left<{\rm C}\psi_{j,+},\psi_{k,-}\right>=\left<{\rm C}\psi_{k,-},\psi_{j,+}\right>$.
We define a linear map $B:L^2({\cal P}_\pm\otimes T\hat{X}\otimes (P \times_G V))\to L^2({\cal P}_\pm\otimes T\hat{X}\otimes (P \times_G V))$:
\begin{align}
B\left\{\sum_ja_j\psi_{j,\pm}\right\}
:=\sum_ja_j{\rm C}\psi_{j,-}.
\end{align}
Since $B$ is a linear operator that changes the basis, we find ${\rm det}(B)=1$.
We finally obtain
\begin{align}
Z_+=&
{\rm Pf}({\cal R}B)
={\rm Pf}({\cal R}).\label{UO-pf-3/2}
\end{align}
We need a regularization to define the partition function (\ref{UO-pf-3/2}) in an infinite value.
Since we cannot introduce a mass term only one of the contributions from $L^2({\cal P}_\pm\otimes T\hat{X}\otimes (P \times_G V))$ \cite{EW},
we add the Rarita-Schwinger field valued in $L^2({\cal P}_-\otimes T\hat{X}\otimes (P \times_G V))$. 
By using $[{\rm C},{\cal R}]=\{\tau,{\rm C}\}=0$, we choose the basis of $L^2({\cal P}_-\otimes T\hat{X}\otimes (P \times_G V))$ as ${\rm C}\psi_{j,+}$. 
Then, we find that the partition function of the Rarita-Schwinger field valued in $L^2({\cal P}_-\otimes T\hat{X}\otimes (P \times_G V))$ is the complex conjugate of $Z_+$ \cite{EW}.
After adding a Pauli-Villars field with mass $M>0$ to the total Lagrangian of the Rarita-Schwinger field defined on $L^2(({\cal P}_+\oplus{\cal P}_-)\otimes T\hat{X}\otimes (P \times_G V))$, we find the regularized partition function of the Rarita-Schwinger field on $L^2({\cal P}_+\otimes T\hat{X}\otimes (P \times_G V))$ is
\begin{align}
Z_+={\rm Pf}\left(\frac{{\cal R}}{{\cal R}+iM}\right).\label{reg-UO3/2}
\end{align}
Here, the operator ${\cal R}/({\cal R}+iM)$ is a linear map on a spinor on $\hat{X}$, and this operator is defined as $({\cal R}/{\cal R}+iM)\psi_j=(\lambda_j/(\lambda_j+iM))\psi_j$.
Since the Pauli-Villars regulator does not break any perturbative gauge-diffeomorphism transformations, this partition function does not have perturbative anomalies.
By a gauge-diffeomorphism transformation $\pi$, the phase of the partition function eq.(\ref{reg-UO3/2}) is changed as $Z_+\to(-1)^{{\rm SF}(f_{R\to R_\pi})}Z_+$, where $f_{R\to R_\pi}$ is defined in eq.(\ref{R-path}).
Furthermore, by using eq.(\ref{flow-equal}), we find the partition function eq.(\ref{reg-UO3/2}) is changed as $Z_+\to(-1)^{{\rm SF}\left(f_{D^{3/2}\to D^{3/2}_\pi}\right)}Z_+$.

We will show that the anomalous phase of the partition function eq.(\ref{pfU-1}) is equivalent to the path integral of a bulk massive spin-$3/2$ Dirac fermion and two bulk massive spin-$1/2$ Dirac bosons.
We consider a $4$-dimensional spin manifold $\hat{Y}$, assuming that $\partial \hat{Y}=\hat{X}$ and $\hat{Y}$ is a cylinder $[0,1]\times \hat{X}$ near the boundary.
We denote $Y=\hat{Y}/\tau_Y$, where $\tau_Y$ is an orientation reversing map that satisfies $\tau_Y^2=1$ and $\tau_Y=\tau$ near the boundary.
Then, $Y$ is a ${\rm Pin}^+(4)$ manifold with a boundary $X$.
A spin-$3/2$ field on $Y$ is an element of the Hilbert space $L^2(\tilde{{\cal P}}_+\otimes T\hat{Y}\otimes (\tilde{P} \times_G V))$, where $\tilde{{\cal P}}_+$ is a ${\rm Pin}^+$ bundle defined in section $2.2$ and $\tilde{P} \times_G V\to \hat{Y}$ is a $G$-bundle on $\hat{Y}$ on a real linear space $V$ which becomes $(P \times_G V)\to \hat{X}$ on the boundary.
We define a new spin-$3/2$ Dirac operator on $Y$ by using the original spin-$3/2$ Dirac operator $i\slashed{D}_{3/2}$ and the Gamma matrices defined in eq.(\ref{4d-gamma}) \cite{EW}:
\begin{align}
i\tilde{\slashed{D}}_{3/2}:=\Gamma^5i\slashed{D}_{3/2}=Ui\slashed{D}_{3/2}U^{-1},\qquad U:=\frac{1-\Gamma_5}{2}.\label{New-UOD}
\end{align}
This new Dirac operator commutes with the orientation reversing map $\tau_Y$:
\begin{align}
[i\tilde{\slashed{D}}_{3/2},\tau_Y]=0.
\end{align}
We define ${\rm C}:=\ast\Gamma^2\Gamma^5$, where $\ast$ is the complex conjugate operator.
We obtain
\begin{align}
{\rm C}^2=-1,\qquad[{\rm C},i\tilde{\slashed{D}}_{3/2}]=0,\qquad [{\rm C},\tau_Y]=0.
\end{align}
Therefore, we can consider a massive spin-$3/2$ Dirac fermion system on $Y$ only the contribution from $\Psi_+\in L^2(\tilde{{\cal P}}_+\otimes T\hat{Y}\otimes (\tilde{P} \times_G V))$:
\begin{align}
S_{3/2}:=\frac{1}{2}\int_{\hat{Y}}d^4x\sqrt{g}
\left\{
\overline{\Psi}_{+\mu}(i\tilde{\slashed{D}}_{3/2}+iM)\Psi^\mu_+
+\overline{{\rm C}\Psi}_{+\mu}(i\tilde{\slashed{D}}_{3/2}+iM){\rm C}\Psi_+^\mu
\right\}.\label{Bulk=UO}
\end{align}
Here, $\overline{\Psi}_+:=\Psi_+^T\tilde{\Gamma}^2$.
We choose the boundary condition:
\begin{align}
L:\:(1-\tilde{\Gamma}^0)\Psi_+^\mu|_{\partial \hat{Y}=\hat{X}}=0.\label{bc-uo}
\end{align}
We also add two massive spin-$1/2$ Dirac bosons $\chi_j\in L^2(\tilde{{\cal P}}_+\otimes (\tilde{P} \times_G V))$ ($j=1,2$):
\begin{align}
S_j=
\frac{1}{2}\int_{\hat{Y}}d^4x\sqrt{g}
\left\{
\overline{\chi}_j(i\tilde{\slashed{D}}_{3/2}+iM)\chi_j
+\overline{{\rm C}\chi}_j(i\tilde{\slashed{D}}_{3/2}+iM){\rm C}\chi_j
\right\}.\label{Bulk1/2=UO}
\end{align}
We should consider only one contribution from each pairs $(\Psi_+,{\rm C}\Psi_+)$ and $(\chi_j,{\rm C}\chi_j)$ ($j=1,2$).
 Then, the $3$-dimensional theory defined by the effective action eq.(\ref{pfU-1}) appears as boundary localized modes.
The path integral of the sum of the action eq.(\ref{Bulk=UO}) and eq.(\ref{Bulk1/2=UO}) is defined by only one of each pairs $(\Psi_+,{\rm C}\Psi_+)$ and $(\chi_j,{\rm C}\chi_j)$ ($j=1,2$) is given as:
\begin{align}
Z^+(Y|L)=\exp(i\pi\eta(i\tilde{\slashed{D}}_{3/2}(Y|{\rm APS})))
\cdot \exp(-i\pi\eta(i\tilde{\slashed{D}}_{1/2}(Y|{\rm APS})))
|Z^+(Y|L)|.\label{UO-RS-PF}
\end{align}
Here, $i\tilde{\slashed{D}}_{3/2}(Y|{\rm APS})$ and $i\tilde{\slashed{D}}_{1/2}(Y|{\rm APS})$ are the spin-$3/2$ and spin-$1/2$ Dirac operators eq.(\ref{New-UOD}) on $L^2(\tilde{{\cal P}}_+\otimes T\hat{Y}\otimes (\tilde{P} \times_G V))$ and eq.(\ref{DOPin4}) on $L^2(\tilde{{\cal P}}_+\otimes (\tilde{P} \times_G V))$ with the APS boundary condition \cite{APS1}.
As in the case of a spin manifold, we find that the change of the path integral by a gauge-diffeomorphism $\pi$ is given as $Z^+(Y|L)\to \exp(i\pi\eta(i\tilde{\slashed{D}}_{3/2}(M_\pi))\cdot \exp(-i\pi\eta(i\tilde{\slashed{D}}_{1/2}(M_\pi)))Z^+(Y|L)$, where $M_\pi$ is the mapping torus of $\pi$.
We also obtain $\exp(i\pi\eta(i\slashed{D}_{3/2}(M_\pi)))=\exp(i\pi\eta(i\tilde{D}_{3/2}(M_\pi)))$ and $\exp(i\pi\eta(i\slashed{D}_{1/2}(M_\pi)))=\exp(i\pi\eta(i\tilde{D}_{1/2}(M_\pi)))$, where $i\slashed{D}_{3/2}(M_\pi)$ and  $i\slashed{D}_{1/2}(M_\pi)$ are the original spin-$3/2$ and spin-$1/2$ Dirac operator on the mapping torus $M_\pi$.
By using the explanation below eq.(\ref{AP-RS}), we find that the global anomaly of the product of the partition function eq.(\ref{reg-UO3/2}) and inverse of the partition function eq.(\ref{pin+pf}) is equivalent to the change of the path integral eq.(\ref{UO-RS-PF}):
\begin{align}
Z^+(Y|L)\to \exp(i\pi\eta(i\tilde{D}_{3/2}(M_\pi)))\cdot
\exp(-i\pi\eta(i\tilde{D}_{1/2}(M_\pi)))
Z^+(Y|L).
\end{align}
This is the anomaly inflow of the Rarita-Schwinger theory on a ${\rm Pin}^+$ manifold.

\section{Global Anomalies}
From the viewpoint of the anomaly inflow, the partition function is defined as a path integral to a bulk system.
However, there is an ambiguity to choose a bulk manifold in eq.(\ref{inflow-phase}) and eq.(\ref{UO-RS-PF}).
We will compare the path integrals eq.(\ref{inflow-phase}) (or eq.(\ref{UO-RS-PF})) on two different bulk spacetimes $Y_1$ with $Y_2$, where $\partial Y_1=\partial Y_2=X$.
This ambiguity is called the Dai-Freed anomaly \cite{Dai-Freed,KY,EW,WY,Garcia}.
In this section, we will first explain that the global anomalies of the Rarita-Schwinger field are classified as classes of the bordism group, and determine the range of global anomalies with some gauge group in the case of a spin manifold.
We will also study the range of global anomalies on a ${\rm Pin}^+$ manifold.

\subsection{Spin Manifold}
Let us consider the Rarita-Schwinger theory on a  $3$-dimensional spin manifold $X$.
By a theorem of the $\eta$-invariant given in \cite{Dai-Freed}, the difference of the phase part of eq.(\ref{inflow-phase}) corresponding to these two different bulk manifolds $Y_1$ and $Y_2$ is as follows \cite{KY}:
\begin{align}
\frac{Z^{3/2}(Y_1|L)}{Z^{3/2}(Y_2|L)}
=&\frac{\exp\left(i\pi\eta(i\slashed{D}_{3/2}(Y_1|{\rm APS}))\right)
\cdot \exp\left(-i\pi\eta(i\slashed{D}_{1/2}(Y_1|{\rm APS}))\right)}{\exp\left(i\pi\eta(i\slashed{D}_{3/2}(Y_2|{\rm APS}))\right)\cdot\exp\left(-i\pi\eta(i\slashed{D}_{1/2}(Y_2|{\rm APS}))\right)}
\notag \\
=&\exp\left(i\pi\eta(i\slashed{D}_{3/2}(Y_1+\overline{Y}_2))\right)
\cdot\exp\left(-i\pi\eta(i\slashed{D}_{1/2}(Y_1+\overline{Y}_2))\right).\label{phase}
\end{align}
Here, $\overline{Y}_2$ the orientation reversing of $Y_2$ and $Y_1+\overline{Y}_2$ the connected sum of $Y_1$ and $\overline{Y}_2$.
Since $Y_1+\overline{Y}_2$ is a $4$-dimensional closed manifold, the difference between two path integrals over two different bulk manifolds eq.(\ref{inflow-phase}) is given as $\exp\left(i\pi\eta(i\slashed{D}_{3/2}({\cal Y}))\right)$, where ${\cal Y}$ is an arbitrary $4$-dimensional spin closed manifold that has a gauge bundle.
By using the APS index theorem \cite{APS1,HS2205}, when exists a $5$-dimensional spin manifold $Z$ that has a gauge bundle and satisfies ${\cal Y}=\partial Z$, the phase difference eq.(\ref{phase}) that corresponds to a closed $4$-dimensional manifold ${\cal Y}$ is trivial.
We denote ${\cal Y}_1\sim{\cal Y}_2$, iff two different $4$-dimensional closed manifold with ${\cal S}$-structures and gauge bundles on ${\cal Y}_1$ and ${\cal Y}_2$ satisfy ${\cal Y}_1+\overline{{\cal Y}}_2=\partial Z$ for a $5$-dimensional spin manifold $Z$ with a gauge bundle.
This relation $\sim$ is an identification called bordism equivalent.
The set of $d$-dimensional closed manifolds with a ${\cal S}$-structure and a $G$-bundle with the bordism equivalence is called the bordism group $\Omega_d^{\cal S}(BG)$.
By using the APS index theorem \cite{APS1}, if there exists a $5$-dimensional manifold $Z$ that satisfies ${\cal Y}_1+\overline{{\cal Y}}_2=\partial Z$, these two closed manifolds ${\cal Y}_1$ and ${\cal Y}_2$ give the same global anomalies.
As explained below eq.(\ref{AP-RS}), we can show that each non-zero eigenvalue of these Dirac operators on ${\cal Y}$ make quartets $(\lambda,\lambda,-\lambda,-\lambda)$.
We also find that the number of positive/negative chiral zero modes are both even (See the explanation below eq.(\ref{AP-RS})).
 Therefore, $\exp\left(i\pi\eta(i\slashed{D}_{3/2}({\cal Y}))\right)=(-1)^{{\cal I}_{3/2}({\cal Y})/2}$ and $\exp\left(-i\pi\eta(i\slashed{D}_{1/2}({\cal Y}))\right)=(-1)^{{\cal I}_{1/2}({\cal Y})/2}$, where ${\cal I}_{3/2}({\cal Y})$ and ${\cal I}_{1/2}({\cal Y})$ are the index of the spin-$3/2$ Dirac operator and the index of the spin-$1/2$ Dirac operator on ${\cal Y}$.
Thus, the phase difference between two different choices of bulk manifolds is classified by the following homomorphism:
\begin{align}
\eta_{3/2}[G]:\:\Omega_4^{\rm Spin}(BG)\to \{\pm 1\};\qquad {\cal Y}\mapsto&\exp(i\pi\eta(i\slashed{D}_{3/2}({\cal Y})))\cdot \exp(-i\pi\eta(i\slashed{D}_{1/2}({\cal Y})))
\notag \\
&=(-1)^{({\cal I}_{3/2}({\cal Y})+{\cal I}_{1/2}({\cal Y}))/2}.
\label{Bor-ano}
\end{align}

These anomalies are canceled by adding a topological term.
Let's consider a counter-term \cite{CH,Bilal}:
\begin{align}
\Delta \Gamma=-\frac{i\pi}{2}\frac{1}{8\pi^2}\int_X\left\{
-{\rm tr}\left(AdA+\frac{2}{3}AAA\right)+\frac{11}{12}{\rm tr}\left(\omega d\omega+\frac{2}{3}\omega\omega\omega\right)
\right\}.\label{counter}
\end{align}
Here, $X$ is an $3$-dimensional spin manifold, $A$ is a gauge field, and $\omega$ is a spin connection defined in eq.(\ref{1.5-order}) without torsion.
If a $4$-dimensional manifold $Y$ satisfies $\partial Y=X$, by using the Stokes theorem, we find
\begin{align}
\Delta\Gamma=-\frac{i\pi}{2}\frac{1}{8\pi^2}\int_Y\left\{
-{\rm tr}(F\wedge F)+\frac{11}{12}{\rm tr}(R\wedge R)
\right\}.
\end{align}
Here, $F$ is the field strength of $A$ and $R$ is the Riemann curvature.
When we add the counter term eq.(\ref{counter}) into the original action eq.(\ref{SG=Action}), we also add the counter term eq.(\ref{SG=Action}) into the bulk action.
Then, the bulk path integral eq.(\ref{inflow-phase}) becomes
\begin{align}
Z^{3/2}(Y|L)=|Z^{3/2}(Y|L)|\exp\left(i\pi\eta(i\slashed{D}_{3/2}(Y|{\rm APS}))-i\pi\eta(i\slashed{D}_{1/2}(Y|{\rm APS}))+\Delta \Gamma\right).\label{inflow-cou}
\end{align}
By using the Atiyah-Singer index theorem \cite{AS1,AS2, HS2205}, we find that this counter-term cancels global anomalies.

\subsection{${\rm Pin}^+$ Unorientable Manifold}
We will consider the case on a ${\rm Pin}^+$ unorientable manifold.
In the same way, the phase difference between two different choices of bulk manifolds are classified by the following homomorphism:
\begin{align}
\tilde{\eta}_{3/2}[G]:\:\Omega_4^{\rm Pin^+}(BG)\to U(1);\qquad {\cal Y}\mapsto\exp(i\pi\eta(i\tilde{\slashed{D}}_{3/2}({\cal Y})))\cdot\exp(-i\pi\eta(i\tilde{\slashed{D}}_{1/2}({\cal Y}))).
\label{UBor-ano}
\end{align}
Therefore, global anomalies are classified as classes of the bordism groups.
The generator of the bordism group $\Omega_4^{\rm Pin^+}(pt)=\mathbb{Z}_{16}$ is the $4$-dimensional real projective space $\mathbb{R}P^4$ \cite{Pin16}.
The $\eta$-invariant of the spin-$1/2$ Dirac operator on $\mathbb{R}P^4$ satisfies \cite{EW,Mod16,TY18}:
\begin{align}
\exp(i\pi\eta(i\tilde{\slashed{D}}_{1/2}(\mathbb{R}P^4))=\exp(\pm i\pi/8).
\end{align}
In the same way as Appendix C of \cite{EW}, we find
\begin{align}
\exp(i\pi\eta(i\tilde{\slashed{D}}_{3/2}(\mathbb{R}P^4)))=\exp(\pm i\pi/2).
\end{align}
Therefore, eq.(\ref{UBor-ano}) is given as follows:
\begin{align}
\tilde{\eta}_{3/2}[pt]:\:\Omega_4^{{\rm Pin}^+}(pt)=\mathbb{Z}_{16}\to U(1),\qquad k\:{\rm mod}\:16\mapsto e^{i\pi 3k/8}.\label{UO-pt}
\end{align}

\section{Conclusions and Discussions}
We considered the anomaly inflow of the Rarita-Schwinger field on $3$-dimensional spin manifold and ${\rm Pin}^+$ manifold.
The global anomalies are obtained as the spectral flow of the Rarita-Schwinger operator.
 We found that the spectral flow of the Rarita-Schwinger operator is equivalent to that of the spin-$3/2$ Dirac operator by using homotopy equivalence of the spectral flow \cite{unbound-F}.
 This result is consistent with the result in \cite{GG=Witten} that the global anomaly of the Rarita-Schwinger field is equivalent to the index of the spin-$3/2$ Dirac operator.
We showed that the partition function of the Rarita-Schwinger field in $3$ dimensions is obtained from the path integral of a massive spin-$3/2$ Dirac fermion on a $4$-dimensional bulk manifold with the chiral boundary condition.
This $4$-dimensional bulk theory gives the anomaly inflow of the Rarita-Schwinger field, where a similar result has been discussed in $11$ dimensions \cite{FH-19}.
We also found that there are no global anomalies Rarita-Schwinger theory on a $3$-dimensional spin manifold.
We determined the global anomaly of the Rarita-Schwinger theory on a ${\rm Pin}^+$ manifold that corresponds to the generator of $\Omega_4^{{\rm Pin}^+}(pt)=\mathbb{Z}_{16}$.

As we discussed in section $3$, we can see that the spectral flow of any operator which is the sum of the Dirac operator and a pseudo-partial differential operator is shown to be equivalent to the spectral flow of the Dirac operator.
Therefore, the anomalous phase of the partition functions of fermion field theories in $d$-dimensions which correspond to those operators are equivalent to the path integral of a massive spin-$3/2$ Dirac fermion on a ($d+1$)-dimensional bulk manifold with the chiral boundary condition.
Since there exists unique invertible topological field theory that corresponds to a bordism invariant in ($d+1$) dimensions \cite{KY-Cob}, it is interesting to explore the set of $d$-dimensional anomalous quantum field theories whose partition functions corresponding to the same bordism invariants in ($d+1$) dimensions.

\section*{Acknoledgements}
The author thanks Yuji Tachikawa for suggesting the topic of this paper and useful comments.
The author also thanks Katsushi Ito for useful comments and discussions, Yosuke Imamura for comments about the supergravity, and Kiyonori Gomi for comments about the proof of the spectral flow in section $3$ and Appendix B and the calculation of the bordism group.
S.K is supported by a postdoctoral Scholarship in DIAS.

\appendix

\section{$1.5$ Order Formalism}
In this section, we will briefly summarize the $1.5$ order formalism.
Let us consider a $3$-dimensional spin manifold.
We denote $x^\mu$ the local coordinate, the metric as $g_{\mu\nu}$.
We denote $e^a(x)=e^a{}_\mu(x)\partial^\mu$ the frame field, where $e^a{}_\mu$ is the vielbein.
The curvature is defined by using the spin connection $\omega_{\mu ab}$ as 
\begin{align}
R_{\mu\nu ab}:=[D_\mu,D_\nu]_{ab}=\partial_\mu\omega_{\nu ab}-\partial_\nu\omega_{\mu ab}+\omega_{\mu ac}\omega_\nu{}^c{}_b-\omega_{\nu ac}\omega_\mu{}^c{}_b.\label{R}
\end{align}
Here, $D_\mu$ is the covariant derivative on the spinor bundle.
The torsion is defined by $T^a:=De^a:=de^a+\omega^a{}_b\wedge e^b=:T_{\mu\nu}{}^adx^\mu\wedge dx^\nu$, where
\begin{align}
T_{\mu\nu}{}^a=\partial_\mu e^a{}_\nu-\partial_\nu e^a{}_\mu+\omega_\mu{}^a{}_b e^b{}_\nu-\omega_\nu{}^a{}_b e^b{}_\mu.\label{T1}
\end{align}
The spin connection is determined by the torsion and vielbein \cite{Freedman-text}:
\begin{align}
\omega_{\mu ab}
=&-\frac{1}{2}e^\nu{}_ae^\rho{}_b(\partial_\mu g_{\nu\rho}+\partial_\nu g_{\mu\rho}-\partial_\rho g_{\mu\nu})
+e^\nu{}_a\partial_\mu e_{b\nu}
\notag \\
&-\frac{1}{2}\left(T_{\mu\nu}{}^ce_{c\rho}-T_{\nu\rho}{}^ce_{c\mu}+T_{\rho\mu}{}^ce_{c\nu}\right) e^\nu{}_ae^\rho{}_b.
\label{spin=con}
\end{align}

Below, we will choose the torsion such that $\delta S^E_{\rm SUGRA}/\delta \omega_{\mu ab}=0$, where $S^E_{\rm SUGRA}$ is the Rarita-Schwinger action defined in eq.(\ref{SG=Action}).
This formalism is called Palatini formalism or $1.5$ order formalism.
The condition $\delta S^E_{\rm SUGRA}/\delta \omega_{\mu ab}=0$ is equivalent to
\begin{align}
-T_{a\rho}{}^\rho e_b^\nu+T_{b\rho}{}^\rho e_a^{\nu}
+T_{ab}{}^\nu
=
\kappa^2\epsilon^{cde}\epsilon_{abf}
e^\mu_ce^\nu_de^\rho_e\psi^\dag_\mu
\sigma^f\psi_\rho.
\label{first-order1}
\end{align}
The following choice of the torsion satisfies eq.(\ref{first-order1}) (See \cite{Freedman-text}):
\begin{align}
T_{ab}{}^\nu=\kappa^2(
\psi_a^\dag\gamma^\nu\psi_b
-\psi^\dag_b\gamma^\nu\psi_a).
\label{torsion}
\end{align}
Substitute eq.(\ref{torsion}) into eq.(\ref{spin=con}), the corresponding spin connection is obtained as
\begin{align}
\omega_{\mu ab}
=&e^\nu{}_a\partial_\mu e_{b\nu}
+\frac{\kappa^2}{2} e_\mu^c\epsilon_{abc}\epsilon^{def}
(\psi_d^\dag\gamma_e\psi_f).
\label{1.5-order}
\end{align}
In this paper, we use this torsion and spin connection.
With the spin connection eq.(\ref{1.5-order}), the action eq.(\ref{SG=Action}) is invariant under eq.(\ref{local=SUSY}).
However, when we choose this spin connection eq.(\ref{1.5-order}), we cannot construct any covariant derivatives on a spinor bundle, because the corresponding covariant derivative cannot be linear.

The covariant derivative $D_\mu$ of the metric $g_{\mu\nu}$ is determined by:
\begin{align}
(D_\mu g)_{\nu\rho}:=\partial_\mu g_{\nu\rho}-\Gamma^\sigma{}_{\mu\nu}g_{\sigma\rho}-\Gamma^\sigma{}_{\mu\rho}g_{\nu\sigma}.
\end{align}
Here, $\Gamma^\sigma{}_{\mu\nu}$ is a connection on the tangent bundle.
We choose the Affine connection:
\begin{align}
\Gamma^\rho{}_{\mu\nu}=\frac{1}{2}g^{\rho\sigma}(\partial_\mu g_{\sigma\nu}+\partial_\nu g_{\mu\sigma}-\partial_\sigma g_{\mu\nu})
+\frac{1}{2}\left(T_{\mu\nu}{}^ae_{a}{}^\rho-T_\nu{}^\rho{}^ae_{a\mu}+T^\rho{}_{\mu}{}^ae_{a\nu}\right).
\label{Affine}
\end{align}
Then, we find
\begin{align}
(D_\mu g)_{\nu\rho}=(D_\mu e)^a{}_\nu=0.
\end{align}

\section{Self-Adjoint Fredholm Operators and Spectral Flow}
We will review the spectral flow of self-adjoint Fredholm operators.
We will first explain the Dirac operator and the Rarita-Schwinger operator are elliptic Fredholm operators, and then define a distance on the space of self-adjoint Fredholm operators and introduce the distance phase on this space.
Then, we review an important feature of the spectral flow, which we use in section $3$.

Let us consider a vector space $V$ with a complex inner product $\left<\quad,\quad\right>:V\times V\to\mathbb{C}$.
We say a norm-space $V$ is a Hilbert space if there exists an element $v\in V$ such that $v_n\to v$ for each Cauchy sequence $\{v_n\}_n$.
For any vector space $V$ with a complex inner product, there exists a vector space ${\cal H}(V)$ with a complex inner product and a norm preserving map $\tau:V\to{\cal H}(V)$ which satisfies the following conditions:
\begin{itemize}
\item[(1)]
${\cal H}(V)$ is Hilbert space.
\item[(2)]
$\tau(V)$ is dense in ${\cal H}(V)$.
\end{itemize}
${\cal H}(V)$ is unique up to isometry, which is the completion of $V$.
In particular, we denote $L^2(E)$ the completion of $C^\infty(E)$, where $C^\infty(E)$ is the vector space of the smooth sections of a smooth vector bundle $E\to X$ on $X$ whose inner product is given by the Lebesgue inner product eq.(\ref{inner-RS}).
By definition, $L^2(E)$ is a Hilbert space.

An operator $T$ on a Hilbert space ${\cal H}$ is Fredholm if ${\rm Ker}(T)$ and ${\rm codim}(T)$ are both finite dimensions, where the codimension of an operator $T$ on a Hilbert space ${\cal H}$ is defined as
\begin{align}
{\rm codim}(T):={\rm dim}({\cal H}/T({\cal H})).
\end{align}
For any self-adjoint Fredholm operator ${\cal F}$ on a Hilbert space ${\cal H}$, the reduced $\eta$-invariant $\eta({\cal F})$  is defined as follows \cite{APS3}: 
\begin{align}
\eta({\cal F}):=&\frac{1}{2}{\rm lim}_{s\to 0}\sum_{j,\;\lambda_j\neq 0}
{\rm sign}(\lambda_j)|\lambda_j|^{-2s}+\frac{1}{2}{\rm dim}\:{\rm ker}({\cal F}).
\label{eta=def}
\end{align}
Here, $\lambda_j\in\mathbb{R}$ are eigenvalues of ${\cal F}$, ${\rm sign}(\lambda)=+1$ ($\lambda>0$) and ${\rm sign}(\lambda)=-1$ ($\lambda<0$).
By the theorem $1.3.2$ of \cite{Spherical}, all elliptic operators on a compact manifold are Fredholm.
To show that the Dirac operator and the Rarita-Schwinger operator are elliptic, we will write down the definition of the $d$-th order pseudo-partial differential operator and the elliptic operator (See  the definition $1.3.3$, definition $1.3.5$, definition $1.3.9$, and definition $1.3.22$ of \cite{Spherical}).
\begin{itemize}
\item
{\bf $d$-th order pseudo-partial differential operator:}
Let $U$ be a bounded open subset of $\mathbb{R}^m$.
Let $p(x,\xi)$ be smooth on $U\times \mathbb{R}^m$ and compact map on $U$, where $(x,\xi)$ is the coordinate on $U\times\mathbb{R}^m$.
Let $S^d(U)$ be the set of smooth maps on $U\times \mathbb{R}^m$ which are compact maps on $U$ that satisfies the following condition for a constant $C(\alpha,\beta,p)$:
\begin{align}
|D^\alpha_xD^\beta_\xi p|\leq C(\alpha,\beta,p)(1+|\xi|)^{d-|\beta|},\:\forall\alpha:=(\alpha,\cdots,\alpha_m),\:\beta:=(\beta_1,\cdots,\beta_m).
\end{align}
Here, $\alpha_j,\beta_j\in\mathbb{N}$, and we use the following notation:
 \begin{align}
 |\beta|:=&\beta_1+\cdots+\beta_m,
 \notag \\
 d_x^\alpha:=&\left(\frac{\partial}{\partial x_1}\right)^{\alpha_1}\cdots\left(\frac{\partial}{\partial x_m}\right)^{\alpha_m},
 \notag \\
 D_x^\alpha:=&(-\sqrt{-1})^{|\alpha|}d_x^\alpha.
 \end{align}
 We find that $S^{d_1}(U)\subset S^{d_2}(U)$ if $d_1<d_2$.
 
 For a $p\in S^d(U)$ and $f\in C^\infty_0(U)$, we define the associated pseudo-differential operator $P$ of order $d$ by
 \begin{align}
 P:\:&C_0^\infty(U)\to C^\infty_0(U);\qquad f\mapsto Pf;
 \notag \\
 Pf(x):=&\int e^{\sqrt{-1}(x-y)\cdot \xi}p(x,\xi)f(y)dyd\xi.\label{pseudo-op}
 \end{align}
 Here, $x,y\in U$ and $\xi\in\mathbb{R}^m$.
 We denote $C^\infty_0(U)$ as the space of smooth function on $U$ with compact support in $U$.
 We denote $\sigma(P):=p$ and call $\sigma(P)$ the symbol of the pseudo-differential operator $Pf(x)$.

\item
{\bf Elliptic Operator:}
 Let $E\to X$ be a vector bundle on a manifold $X$ and we denote $P$ a pseudo-differential of order $d$ on $C^\infty(V)$ with leading symbol $p$.
We say that $P$ is elliptic if there exists $\epsilon>0$ such that
\begin{align}
\sqrt{\left|\left<p(x,\xi)\cdot f,\:p(x,\xi)\cdot f\right>\right|}\geq \epsilon\cdot\sqrt{|\left<f,f\right>|}\cdot|\xi|^d\:{\rm for}\:|\xi|\geq \epsilon^{-1}\:{\rm and}\:\forall f\in C^\infty(E).\label{elliptic}
\end{align}
Here, $\left<\quad,\quad\right>$ is the Lebesgue inner product.
\footnote{
Since the completion is a norm-preserving map, we can consider $f\in L^2(E)$ instead of $f\in C^\infty(E)$.
}

\end{itemize}
To see that the Rarita-Schwinger operator eq.(\ref{RS-op}) is elliptic, we use the following theorem:
\begin{itemize}
\item
Theorem (B1): 
For an elliptic operator $P$ and a pseudo-differential operator of order $d-1$ on $C^\infty(V)$ $q$, $P+Q$ is elliptic (See the explanation in the definition $1.3.22$ in \cite{Spherical}).
\end{itemize}

Now let us show the following fact by using theorem (B1):
\begin{itemize}
\item
Lemma (B2):
The Rarita-Schwinger operator defined in eq.(\ref{RS-op}) and spin-$3/2$ Dirac operator are both elliptic.
\end{itemize}
\begin{proof}
We will first confirm that the spin-$3/2$ Dirac operator $\slashed{D}_{3/2}$ is elliptic.
$\slashed{D}_{3/2}$ is a pseudo-differential operator of order-$1$ whose symbol $\sigma(\slashed{D}_{3/2})$ is
\begin{align}
p_D(x,\xi):=\sigma(\slashed{D}_{3/2})=\gamma^\mu\xi_\mu+\gamma^\mu(A_\mu+\Gamma_\mu+\omega_\mu),\qquad \xi\in\mathbb{R}^3.\label{sym-dirac}
\end{align}
Here, $\Gamma_\mu$ is the Affine connection and $\omega_\mu$ the spin connection.
We find that eq.(\ref{sym-dirac}) satisifies eq.(\ref{elliptic}) with $d=1$:
\begin{align}
\left<p_D(x,\xi)\cdot \psi,p_D(x,\xi)\cdot \psi\right>
=&
\int d^3x |\xi+A(x)+\omega(x)|^2\psi^{\mu\dag}\psi_\mu
+\Gamma_\nu{}^\mu{}_\rho\Gamma^\nu{}_{\mu\kappa}\psi^{\dag \rho}\psi^\kappa
\notag \\
\geq&
\int d^3x\left\{|\xi|^2-|A(x)+\omega(x)|^2
-|\Gamma_\nu{}^\mu{}_\rho\Gamma^\nu{}_{\mu\rho}|\right\}\psi^{\dag\sigma}\psi_\sigma
\notag \\
\geq&
\left\{|\xi|^2-|M|^2\right\}\left<\psi,\psi\right>.
\end{align}
Here, $\left<\quad,\quad\right>$ is the Lebesgue inner product eq.(\ref{inner-RS}), and $|M|^2$ is the maximal value of $|A(x)+\omega(x)|^2
+|\Gamma_\nu{}^\mu{}_\rho\Gamma^\nu{}_{\mu\rho}|$ as a function of $x$.
$\psi$ is a section of the tensor product of the tangent bundle and a spinor bundle on a spin manifold $X$.
Therefore, the symbol eq.(\ref{sym-dirac}) satisfies eq.(\ref{elliptic}),
and we find $i\slashed{D}_{3/2}$ is elliptic.
In the same way, the spin-$1/2$ Dirac operator is elliptic.
Next, we will consider the Rarita-Schwinger operator.
The Rarita-Schwinger operator ${\cal R}$ defined in eq.(\ref{RS-op}) can be written as ${\cal R}=i\slashed{D}_{3/2}+Q$, where 
\begin{align}
(Q\psi)^\mu:=-\gamma^\mu\gamma^\nu\psi^\rho(i\partial_\nu\gamma_\rho).
\label{spin-dirac}
\end{align}
This operator $Q$ is a pseudo-differential operator and the corresponding symbol is $\sigma(Q):=Q$ of order zero.
Therefore, by using theorem (B1), we find that the Rarita-Schwinger operator eq.(\ref{RS-op}) is elliptic.
\end{proof}

Let us introduce a phase on the set of the self-adjoint Fredholm operators.
Consider a manifold $X$ and a vector bundle $E\to X$ on $X$.
We define
\begin{align}
{\cal CF}^{sa}(E):=\{T:{\rm closed}\:{\rm densely}\:{\rm defined}\:{\rm Self}\:{\rm adjoint}\:{\rm Fredholm}\:{\rm operator}\:{\rm on}\:L^2(E)\}.
\label{fredholm}
\end{align}
We said an operator $T:L^2(E)\to L^2(E)$ is densely defined if and only if the domain of the operator $D(T)$ is a dense subset of $L^2(E)$, and we said an operator $T:L^2(E)\to L^2(E)$ is closed if and only if for any arbitrary converges low $\{x_n\}_{n\in\mathbb{N}}\subset D(T)$ that satisfies $x_n\to x\in L^2(E)$ ($n\to\infty$) and $Tx_n\to y\in L^2(E)$, $x\in D(T)$ and $Tx=y$.
We define the gap metric on ${\cal CF}^{sa}(E)$ (See section 1.1 in \cite{unbound-F}), and the gap metric induce the following distance function on ${\cal CF}^{sa}(E)$ (See Theorem 1.1 in \cite{unbound-F}):
\begin{align}
d(T_1,T_2):=||\kappa(T_1)-\kappa(T_2)||,\label{dis}
\end{align}
where the Cayley-transform $\kappa:{\cal CF}^{sa}(E)\to\{U:{\rm unitary}|U-I\:{\rm is}\:{\rm injective}\}$ is defined as
\begin{align}
\kappa(T):=(T-i)(T+i)^{-1}=I-2i(T+i)^{-1}.\label{spec-uni}
\end{align}
The image $\kappa(T)$ is a bounded operator \cite{Gap-lec}, which means that
\begin{align}
||\kappa(T)u||\leq c||u||,\qquad u\in L^2(E).\label{dis-2}
\end{align}
 for a constant $c>0$.
 Then, we define the resolvent norm \cite{Gap-lec}:
 \begin{align}
 ||\kappa(T)||:={\rm sup}_{u\neq 0}\frac{||\kappa(T)u||}{||u||}.\label{dis-3}
 \end{align}
Then, we can define a distance phase ${\cal O}$ on ${\cal CF}^{sa}(E)$ by this distance function eq.(\ref{dis}):
\begin{align}
{\cal O}(E):=\{U_\epsilon(T)\subset {\cal CF}^{sa}(E)|\:T\in{\cal CF}^{sa}(E),\:U_\epsilon(T):=\{T'\in{\cal CF}^{sa}(E)|\;d(T,T')\leq \epsilon\}\}.\label{OS-gap}
\end{align}
With this gap phase, the set ${\cal CF}^{sa}(E)$ is path connected (See Theorem 1.10 of \cite{unbound-F}).
The spectral flow is as follows (See definition 2.12 and Lemma 2.9 of \cite{unbound-F}):
\begin{itemize}
\item
Let $f:[0,1]\to {\cal CF}^{sa}(E)$ be a continuous path.
By the proposition 2.10 in \cite{unbound-F}, we can choose a partition, $\{0=t_0<t_1<\cdots<t_n=1\}$ of the interval $[0,1]$ and a real number $\epsilon_j>0$, $j=1,\cdots, n$ such that for each $j=1,\cdots,n$ the following map is continuous and finite rank on the interval $[t_{j-1},t_j]$:
\begin{align}
E_j:\:[t_{j-1},t_j]\to\{{\rm Bounded}\:{\rm linear}\:{\rm operators}\};\qquad
t\mapsto \frac{1}{2\pi i}\int_{\Gamma_j}(\lambda-f(t))^{-1}d\lambda.
\label{sflow-1}
\end{align}
Here, we denote $\Gamma_j$ the circle of radius $(\epsilon_j-\epsilon_{j-1})$ and centre $(\epsilon_j+\epsilon_{j-1})/2$.
Then, we define the spectral flow of $f:[0,1]\to{\cal CF}^{sa}(E)$ to be
\begin{align}
{\rm SF}(f):=\sum_{j=1}^n\left\{{\rm dim}(E_j^\geq(t_j))-{\rm dim}(E_j^{\geq}(t_{j-1}))\right\}.
\label{sflow-2}
\end{align}
Here, we denote $E_j^\geq (t)$ the space of positive eigenmodes of the operator $E_j(t)$.
\end{itemize}
By proposition 2.13 in \cite{unbound-F}, this definition of the spectral flow does not depend on the partition $\{0=t_0<t_1<\cdots<t_n=1\}$ of the interval $[0,1]$ and a real number $\epsilon_j>0$, $j=1,\cdots, n$.
The spectral flow only depends on the map $f:[0,1]\to {\cal CF}^{sa}(E)$.

\begin{itemize}
\item
Lemma (B3):
The spectral flow is homotopy equivalence (See prop 2.16 in \cite{unbound-F}).
\end{itemize}

We will finally explain the change of the reduced $\eta$-invariant of a Fredholm operator by a gauge-diffeomorphism transformation captured by the spectral flow.
By the theorem 1.10 of \cite{unbound-F}, the space ${\cal CF}^{sa}(E)$ is a path connected with the gap phase.
Therefore, there exists at least one path that connects two different points on ${\cal CF}^{sa}(E)$.
We consider a continuous path $f:[0,1]\to{\cal CF}^{sa}(E)$ that connect two different elliptic self-adjoint operators ${\cal F}_1$ and ${\cal F}_2$ that have the same spectral.
If we can choose the path $f$ on the set of elliptic self-adjoint operators, we obtain the difference of the reduced $\eta$-invariant of these operators as follows (See proposition $2.12$, eq.($7.1$), p.$79$, and p.$94$ of \cite{APS3}, where the reduced $\eta$-invariant is dented as $\xi$):
\begin{align}
\eta({\cal F}_2)-\eta({\cal F}_1)={\rm SF}(f).\label{dif-eta-1}
\end{align}
Here, ${\rm SF}(f)$ is the spectral flow of $f$ defined in eq.(\ref{sflow-2}).

\end{document}